\documentclass[%
 reprint,
 amsmath,amssymb,
 aps,
]{revtex4-2}

\usepackage{siunitx}
\usepackage{graphicx}
\usepackage{dcolumn}
\usepackage{bm}

\newcommand{\ecmtwo}{e\textsuperscript{$-$}/cm\textsuperscript{2}}
\newcommand{\vn}{V\textsuperscript{0}}
\newcommand{\nvm}{NV$^{-}$}
\newcommand{\nvn}{NV\textsuperscript{0}}

\newcommand{\vm}{V$^{-}$}

\newcommand{\Ns}{N\textsubscript{s}}

\newcommand{\Cione}{C\textsubscript{i$\langle 001 \rangle$}}
\newcommand{\Vtwo}{V\textsubscript{2}}
\newcommand{\Vthree}{V\textsubscript{3}}
\newcommand{\Vfour}{V\textsubscript{4}}
\newcommand{\nvv}{NV\textsubscript{2}}
\newcommand{\nnv}{N\textsubscript{2}V}

\begin{document}

\preprint{APS/123-QED}

\title{200 keV energy electron irradiation of single crystal diamond: Quantification of
vacancy and nitrogen-vacancy production}

\author{Chloe C. Newsom}
\author{Ben L. Green}
\author{Mark E. Newton}
\email{M.E.Newton@warwick.ac.uk}
\affiliation{%
 Department of Physics University of Warwick, Coventry, CV4 7AL, UK
}%

\author{Lillian B. Hughes}
\affiliation{
 Department of Materials Science University of California, Santa Barbara 4105 CA 93106-9530 USA
}%

\author{Ania C. Bleszynski Jayich}
\affiliation{%
 Department of Physics University of California, Santa Barbara 4105 CA 93106-9530 USA
}%


\date{\today}

\begin{abstract}
Electron irradiation and annealing treatments are a method of colour centre/defect creation in diamond. The depth profile of defects created by low energy 200~keV electrons in single crystal high purity (type II) electronic grade diamond grown via chemical vapour deposition has been investigated. The depth profile of monovacancies created was found, using photoluminescence (PL), to decay exponentially, with decay length $12\pm1~\unit{\um}$ and production rate of $(1.1\pm0.2)\times 10 ^{-1}$~V/e\textsuperscript{-}/cm at the surface. 
The depth distribution of the neutral and negatively charged nitrogen-vacancy (NV\textsuperscript{0/-}) centre, and the 733~nm zero phonon line defect formed during an isochronal annealing study have also been investigated by PL. The observed NV profiles, which do not match the vacancy profiles, can be qualitatively explained in terms of a simple model that includes the formation of vacancy clusters and the nitrogen-divacancy (\nvv ) defect. The production of (NV\textsuperscript{0/-}) has been assumed to be nitrogen limited, but this paper has shown that this is not the case, with NVs lost to the production of \nvv\ when the concentration of vacancies greatly exceeds that of substitutional nitrogen.

\end{abstract}

\maketitle


\section{Introduction}
The nitrogen-vacancy (NV) centre in diamond \cite{Doherty2013TheDiamond} is a point defect that has received significant interest in recent years.
The negatively charged state of the centre (\nvm) exhibits long spin coherence times and due to its sensitivity to electric and magnetic fields \cite{Wolf2015SubpicoteslaMagnetometry} and intrinsic room-temperature optical spin initialisation and readout, it has many potential applications such as sensing \cite{Dolde2011Electric-fieldSpins, Schirhagl2014Nitrogen-vacancyBiology, Kucsko2013Nanometre-scaleCell}, nanophotonics \cite{Riedel2017DeterministicDiamond, Aharonovich2014DiamondNanophotonics}, quantum memory \cite{Poem2015BroadbandDiamond} and quantum information processing \cite{Benjamin2009ProspectsSpins, Nickerson2014FreelyLinks}.

Both single NV centres and ensembles of centres have been utilised across these technology platforms and hence the reliable creation of NV centres on demand is an important research field \cite{Eichhorn2019OptimizingSensing, Chen2019LaserYield, Edmonds2021CharacterisationApplications}. 



The neutral \nvn\ has an optical zero phonon line (ZPL) at 575~nm (2.156~eV). \nvn\ can also accept an electron from donors in the diamond forming \nvm , which possesses an optical absorption and luminescence ZPL at 637~nm (1.945~eV). \nvm\ can be converted to \nvn\ via photoionisation using light \cite{Beha2012OptimumDiamond, Siyushev2013OpticallyTemperatures, Razinkovas2021PhotoionizationCalculations}. However at room temperature $>95$\% of the of the total absorption and emission of both centres is via the broad vibronic sideband. 

The precise positioning of these functional defects, such as single \nvm\ centres, and the optimisation of \nvm\ ensembles, whilst minimising the concentration of unwanted defects that degrade the useful \nvm\ properties is desirable \cite{Smith2019ColourTechnologies}. To achieve this it is important to understand diffusion in the lattice, and the formation/destruction kinetics of related defects. 
Intrinsic defects such as interstitials and vacancies are  known to assist the diffusion of carbon atoms and impurities \cite{Orwa2012AnDiamond, Onoda2017DiffusionDiamond}. Therefore, the understanding of these intrinsic defects is key in the pursuit of deterministic defect creation.

In this work we analyse the depth profile of the damage created by low energy (200~keV) electron irradiation in high purity, ``electronic grade'' diamond ($<$ 5~ppb nitrogen, $<$ 1~ppb boron). An isochronal annealing study has been carried out to measure the depth profile of the defects created, focusing on NV\textsuperscript{$0/-$} centres and a centre with a zero phonon line (ZPL) at 733~nm  believed to be a multivacancy complex \cite{Steeds2014AnnealingSamples}. It is an interesting finding that the peak of the NV emission is at a depth of 60~\unit{\um} below the surface, where the vacancy emission peaks.

\subsection{Electron irradiation in diamond}

The most common method of vacancy and carbon interstitial creation is irradiation or ion implantation. Irradiation can be carried out using neutrons, protons, gamma rays and electrons, all which differ in the damage profiles they cause. Ion implantation of non carbon elements has the added effect of introducing impurities into the lattice, in the form of the ions used \cite{Pezzagna2011CreationRemarks}. 
Irradiation in conjunction with annealing treatments can be used to investigate diffusion, along with the mechanisms through which vacancies and interstitials are lost \cite{McLellan2016PatternedTechnique, Eichhorn2019OptimizingSensing, Chen2017LaserDiamond}. Electon irradiation is particularly attractive because of the ability to control the vacancy production [ref]. The quantification of the vacancy production efficiency of irradiation methods is therefore important for the formation of defects.

The threshold energy of electrons required to create damage in diamond has been measured to be $145\pm15~$keV for a (100) incident direction, consistent with a displacement energy for a lattice carbon atom of $30\pm 4$~eV \cite{McLellan2016PatternedTechnique}. 
Monte Carlo simulations have been used to predict the number of vacancies created per micrometer per depth created for 1~MeV, 2~MeV and 5~MeV over a depth of 8,000~\unit{\um} and the resultant profiles demonstrated constant vacancy production profiles with sharp cut offs at a maximum depth \cite{Campbell2000RadiationIrradiation}. 
CASINO simulations of the depth profile of lower energy electrons (200~keV) have predicted a gradual decrease in vacancy production over a depth of 38~\unit{\um}, beyond which the incident electrons are no longer able to generate vacancies \cite{Losero2023CreationIrradiation}. Previous experimental measurements of the depth profiles of vacancies created by 300~keV electrons using a transmission electron microscope have been observed to decrease gradually to a depth of 140~\unit{\um} \cite{Wang2017AnnealingIrradiation}.

\begin{figure}
\includegraphics[width=0.98\columnwidth]{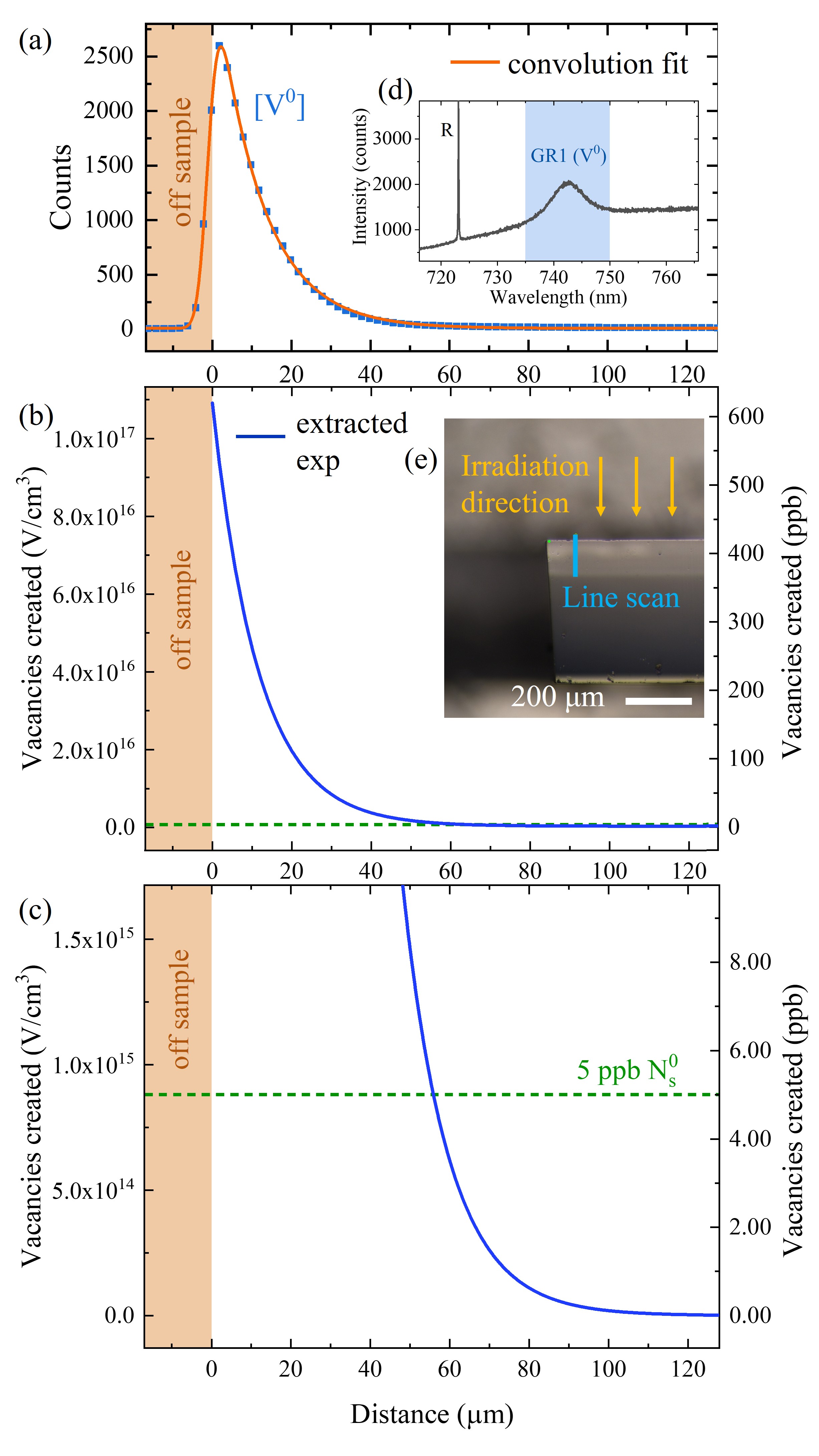}
\caption{\label{Vdepth} (a) Room temperature PL line scan showing the spatial depth profile of GR1 sideband as irradiated, fit with a convolution of a Gaussian and peak and an exponential decay. 660~nm excitation used with a 0.9~NA $\times$100 objective, resulting in a 1~\unit{\um} spot size. (b) Plot of the exponential extracted from the convolution fit in (a). Area under the exponential has been normalised to the concentration determined by absorption measurements to calculate \vn\ concentration with depth. Dotted line shows estimated nitrogen concentration present in the sample. (c) Magnified region of (b). (d) Example room temperature GR1 spectra. Shaded region indicates integrated area. (e) White light image of the location of the line scan marked in blue. }
\end{figure}

\section{Experimental detail}
\subsection{\label{sec:Sample}Sample preperation}
The sample studied here is an (001) oriented $2.09 \times 2.09 \times 0.52$~mm electronic grade plate grown via chemical vapour deposition (CVD) by Element Six \cite{ElementSixTechnologies2023Https://e6cvd.com/us/application/quantum-radiation/el-sc-plate-2-0x2-0x0-5mm.html}. It is specified to contain $<1$~ppb boron and $<5$~ppb nitrogen impurities.

The sample was irradiated using a “PCT Ebeam and Integration” CE60120 Ebeam lamp. 200 keV electrons were used with an estimated dose of $10^{19}$~\ecmtwo\ perpendicular to the 2.09~x~2.09~mm face. The electrons travelled through air before arrival at the sample surface, however the path length was minimised to reduce energy loss (approx. 32~mm). A comparison of the depth profile with that of an irradiation carried out in a vacuum by a 200~keV transmission electron microscope indicated that the energy loss was negligible.
After irradiation the 2.09 x 0.52~mm laser cut sides were polished to allow photoluminescence (PL) measurements of the depth profile of luminescent defects from the polished sides.

\subsection{\label{sec:Methods}Methods}
Optical absorption measurements were made post-irradiation, using a Perkin Elmer Lambda 1050 spectrometer through the large face in the same direction as irradiation, with the sample held at 77K.

The sample was annealed in a tube furnace for an hour at 400\textdegree C, 600\textdegree C, 800\textdegree C, 1000\textdegree C and 1200\textcelsius. At each annealing stage the sample was loaded at the stated annealing temperature, a process with takes approx. 15 minutes. At higher temperatures, the change in concentration of defects can be significant even for short loading times. 

Confocal PL line scans were were taken both at room temperature and at 77~K in a liquid nitrogen cryostat across the 2.09 x 0.52~mm side to measure the depth profile of the defects after each annealing stage. Spectra were taken using 660~nm, 514~nm and 488~nm excitation on two different spectrometers, a Horiba LabRam HR Evolution and a Renishaw inVia Raman Microscope.

\section{Results}

\subsection{\label{sec:Vacancy}Vacancy damage profile}


\begin{figure}
\includegraphics[width=\columnwidth]{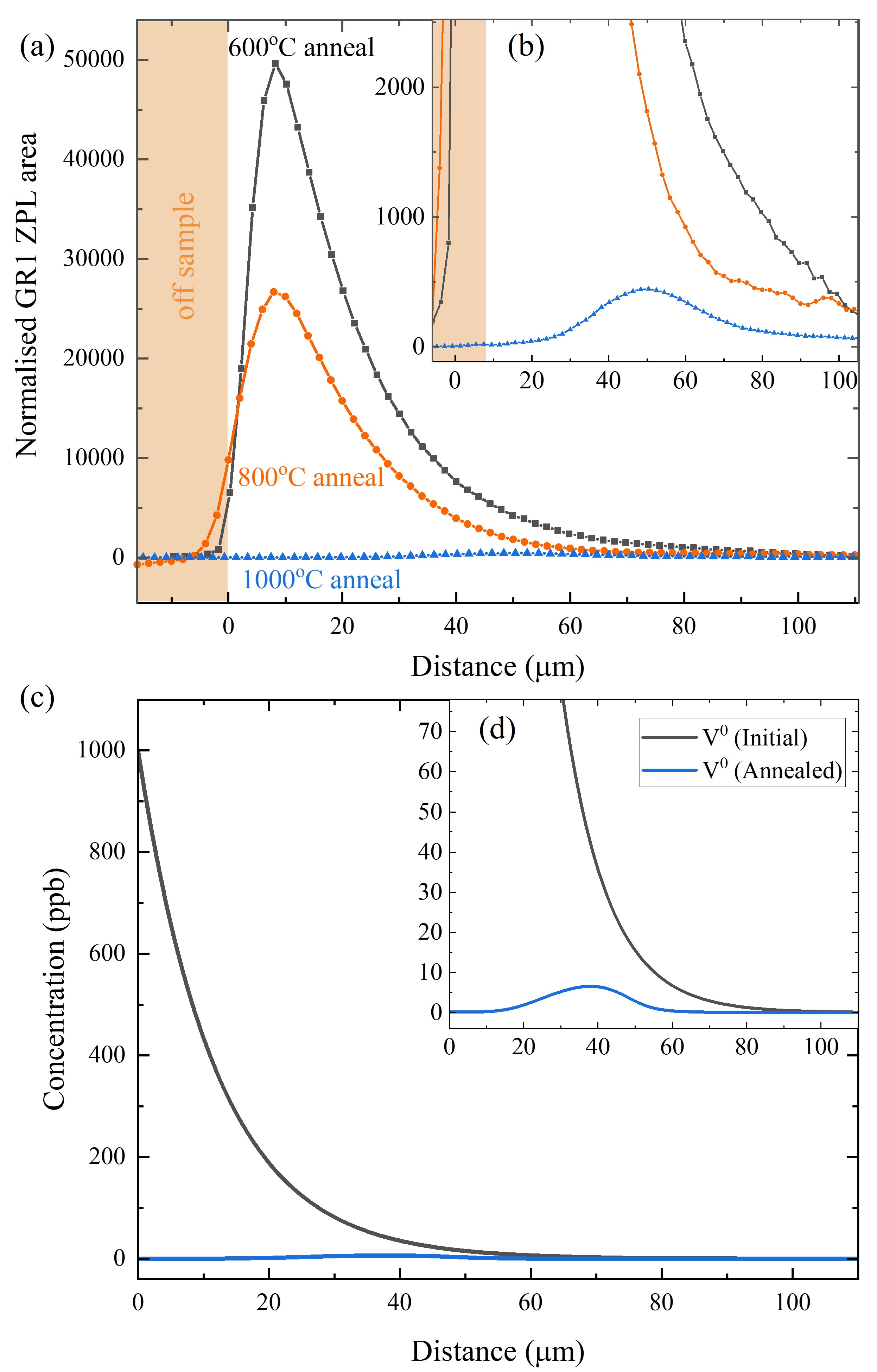}
\caption{\label{VdepthAnneal} (a) Cold PL line scans showing the spatial depth profile of GR1 ZPL after a one hour anneal at 600\textcelsius\ and 800\textcelsius\ and 1000\textcelsius . Spectra were taken at 77~K and have been normalised by the Raman line. 660~nm excitation used with a 0.5~NA $\times$50LWD objective. Units are counts$\cdot$nm. (b) Magnified region of (a) for clarity.  (c) Simulated vacancy concentration before and after a 1 hour anneal at $T_1 = 1000$\textcelsius . (d) Magnified region of (c).}
\end{figure}

After irradiation, absorption measurements were performed to quantify the irradiation damage. Only the neutral charge state of the vacancy \vn , was visible in the absorption spectra with a ZPL at 741~nm (known as the GR1 peak). No signal was observed from the negatively charged vacancy \vm\ in absorption, which has a ZPL at 393.5~nm (ND1). Given the sample contains $<$5~ppb N\textsubscript{s}\textsuperscript{0} the majority of the vacancies are expected to be in the neutral charge state after irradiation \cite{Davies1992Vacancy-relatedDiamond}.
The average number of vacancies per unit volume was measured to be $(2.5\pm 0.5)\times10^{15}$~V/cm$^3$ across the whole diamond thickness of 0.52~mm, equivalent to 14$\pm 3$~ppb, if the vacancies were homogeneously distributed through depth. 
A calibration constant of $(1.2\pm 3) \times 10^{-16}$~meV cm$^2$ \cite{Twitchen1999CorrelationEPR} was used to convert the area under the GR1 ZPL to a concentration.

To determine the depth profile of the \vn\ PL after irradiation, a line scan of the GR1 sideband at room temperature was taken across the polished side of the sample with a 2~\unit{\um} step size, as indicated in fig. \ref{Vdepth}(e). The laser was focused 10~\unit{\um} below the surface of the side. The measured depth profile is a convolution between the actual defect concentration profile and the response function of the spectrometer.
In order to analyse the data some assumptions have been made. An exponential decay has been assumed for the \vn\ concentration as a function of depth from the irradiated surface, $h(x)$, supported by results reported in previous studies \cite{Wang2017AnnealingIrradiation} and a Gaussian function with a width of 2.1~\unit{\um} was used to model the spectrometer response function. 
The resultant convolution, detailed in the supporting information, was fit to the GR1 depth profile (fig. \ref{Vdepth}(a)), allowing the exponential decay constant to be determined as $t_0 = 12\pm1~$\unit{\um}. 



Having determined the decay constant of the GR1 PL depth profile, we can now use the PL in conjunction with the absorption measurements to calculate the spatial distribution of vacancies per unit volume with depth (assuming the PL intensity of the GR1 sideband is proportional to \vn\ concentration). By integrating over the distribution measured using PL and normalising the area to the absorption data, the PL depth profile can be converted into a quantitative vacancy depth profile, as shown in fig. \ref{Vdepth}(b). 

The \vn\ profile shown in fig. \ref{Vdepth}(b) is given by
\begin{equation}
    [\vn ] = (1.1\pm 0.2 \times 10^{17}) \exp \left( {-\frac{x}{12\pm 1}} \right)/\text{cm}^{-3}.
    \label{eq: UCSBVConcEq}
\end{equation}
Equation \ref{eq: UCSBVConcEq} can be used to derive the \vn\ concentration at the surface: $(1.1\pm 0.2) \times 10^{17}$~V/cm$^3$ or $600 \pm 100$~ppb.

To study the annealing of vacancies as a function of temperature, line scans of the GR1 PL depth profile from the polished side of the sample were measured at each stage of an isochronal annealing study. Measurements made during the annealing study were carried out at liquid nitrogen temperatures, and therefore the defect ZPLs were fitted instead of the sidebands. 
Due to the cryostat used, measurements were made using a $\times$50 long working distance objective with a numerical aperture of 0.5, as opposed to the $\times$100 objective used to measure the profile in fig. \ref{Vdepth}. The spectrometer response function of these measurements is therefore different; further discussion of this is given in \cite{Newsom2023SpectroscopicDiamond}.
No change was observed to the \vn\ GR1 PL intensity depth profile after the 400\textdegree C and 600\textdegree C anneals. The \vn\ PL depth profile is decreased by the same factor across the entire depth profile by the 800\textdegree C anneal, shown in fig. \ref{VdepthAnneal}(a).
This suggests that the fractional vacancy concentration loss is uniform throughout the sample. After the 1000\textdegree C anneal however, the profile has changed shape, fig. \ref{VdepthAnneal}(b), with the highest \vn\ concentration shifting to a depth of $49\pm2~$\unit{\um}. There was no GR1 visible after the anneal at 1200\textdegree C, indicating the vacancies have formed complexes, recombined with interstitials, or lost to extended defects or surfaces.

\subsection{\label{sec:NV} Nitrogen-vacancy creation}


\begin{figure}
\includegraphics[width=\columnwidth]{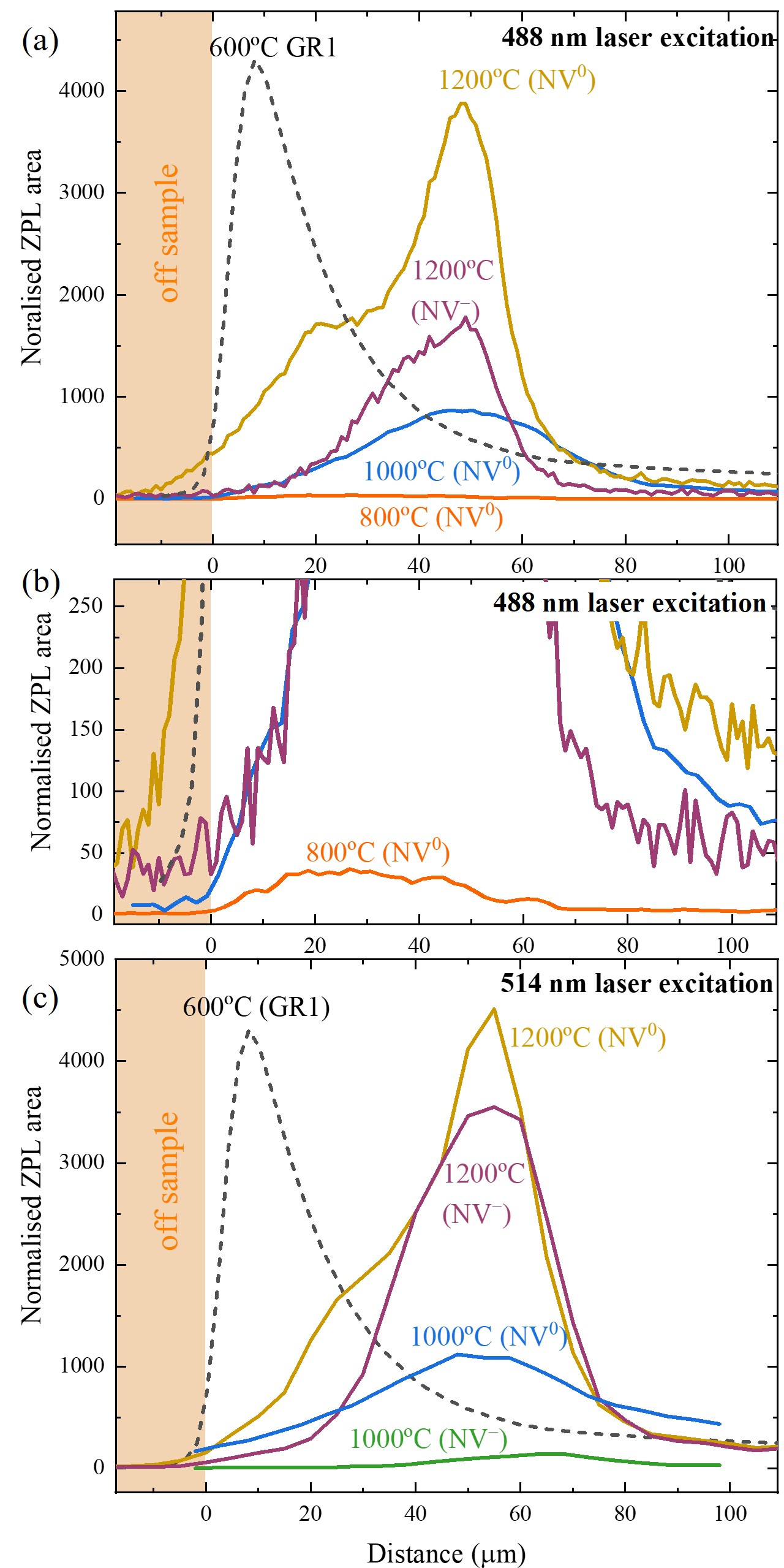}
\caption{\label{fig:NVanneal} PL line scans at 77~K showing the spatial depth profile of the area of the \nvn\ and \nvm\ ZPL after a one hour anneal at each temperature.  Units are counts$\cdot$nm. (a) Line scans taken with 488~nm laser excitation normalised to the Raman line. (b) Magnified region of (a) for clarity. (c) Line scans taken with 514~nm laser excitation normalised to the second order Raman line. }
\end{figure}

Emission from \nvn /\nvm centres is detectable after the 800\textdegree C anneal (fig. \ref{fig:NVanneal}). Measurements of the \nvn\ and \nvm\ ZPL PL depth profiles were made using both 488~nm and a 514~nm laser excitation. The 488~nm excitation causes partial photoionisation \nvm $\rightarrow$ \nvn , whereas the 514~nm laser allows the difference in the distribution of the \nvn\ and \nvm\ charge states to be probed without significant perturbation of the charge state distribution. 

After the 800\textcelsius\ anneal the distribution of \nvn\ was observed to be approximately Gaussian, assuming ZPL PL is proportional to concentration, with a peak position at a depth of $30\pm 5~$\unit{\um} below the irradiation surface, fig. \ref{fig:NVanneal}(b). After the 1000\textcelsius\ anneal an increase in \nvn\ centres is observed across the whole depth profile, as seen with the 488~nm laser, and the peak of the depth profile is shifted to $48\pm5~$\unit{\um}, (fig. \ref{fig:NVanneal}(a)). The measurements taken with the 514~nm laser show a distinct difference in the profiles of the \nvn\ and \nvm\ centres (fig. \ref{fig:NVanneal}(c)). It is seen that the \nvm\ centres are present deeper into the diamond, with a profile that peaks at $64\pm5~$\unit{\um}, as opposed to the \nvn\ profile which peaks at $54\pm5~$\unit{\um}. 

After the 1200\textdegree C anneal, another increase in both \nvn\ and \nvm\ PL across the whole depth profile is observed with both lasers, however the shape of the \nvn\ profile changes significantly, developing a shoulder in the shallow depth region. Signal from \nvm\ centres is also visible with the 488~nm laser, where previously it was not. The \nvn\ profile measured with 488~nm excitation peaks at $49\pm5~$\unit{\um}, while the \nvm\ profile peaks at $51\pm5~$\unit{\um}. The \nvn\ profile measured with 514~nm excitation peaks at $55\pm5~$\unit{\um}, while the \nvm\ profile peaks at $60\pm5~$\unit{\um}. 



\subsection{\label{sec:Other}The 733~nm ZPL centre}

\begin{figure}
\includegraphics[width=\columnwidth]{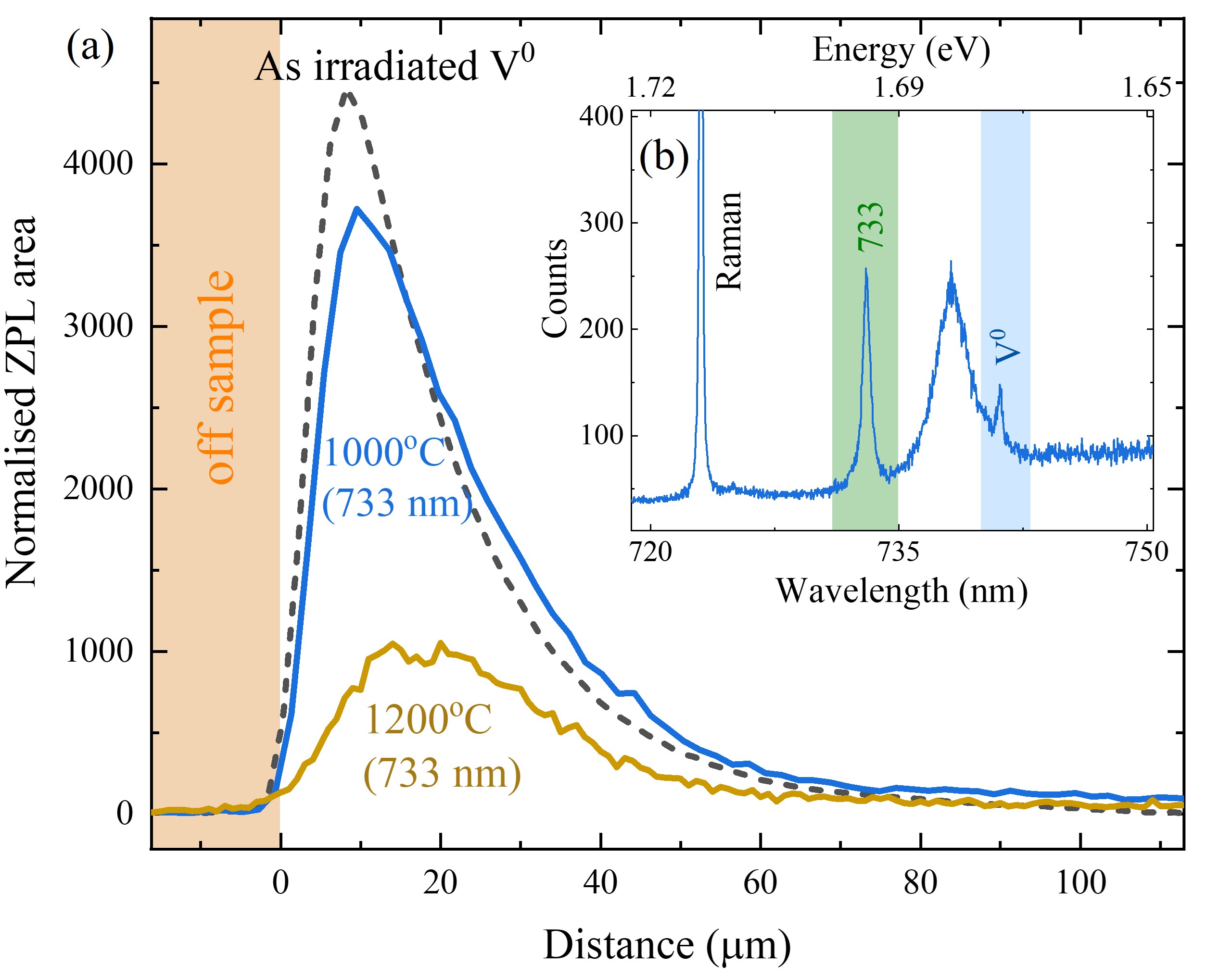}
\caption{\label{733depthAnneal} (a) PL line scans taken at 77K showing the spatial depth profile of the area of the 733~nm ZPL multi-vacancy defect. Dotted line shows scaled \vn\ for comparison. 660~nm laser excitation used.  Units are counts$\cdot$nm. (b) Example spectra of the 733~nm ZPL and sideband, taken at 77K, taken after the 1000\textcelsius\ \ anneal.}
\end{figure}

The 733~nm ZPL has in recent years been attributed to a centre consisting of multiple vacancies \cite{Kiflawi1997TheCrystals, Zaitsev2017OpticalDiamond, Mills2022}. It has been observed in high purity type IIa diamond irradiated and annealed to temperatures above 700\textcelsius\ \cite{Zaitsev2017OpticalDiamond, Steeds2014AnnealingSamples, Kiflawi1997TheCrystals, Wang2017AnnealingIrradiation, Mills2022}. 

In this work it is observed to anneal in after the 1000\textdegree C anneal and can be excited with the 488~nm, 514~nm and 660~nm lasers, though it is strongest with 660~nm excitation. The 733~nm PL anneals in with a depth profile the same as the initial GR1 PL profile, shown in fig. \ref{733depthAnneal}(a).  After the 1200\textdegree C anneal, it can be seen to drop in intensity across the entire profile, but keeps the same profile shape and is not detectable after the 1200\textcelsius\ anneal. An example spectra of the feature is displayed in fig. \ref{733depthAnneal}(b).

This work indicates there is a very narrow temperature window in which the 733~nm centre is detectable; between 1000\textcelsius\ and 1200\textcelsius. 
Its depth profile at both annealing stages distinctly matches that of the initial irradiation corroborating its dependence on a high vacancy concentration for formation.

A number of other spectral features were measured during the annealing process and a list is given in \cite{Newsom2023SpectroscopicDiamond}.

\section{\label{sec:Discussion}Discussion}
\subsection{Intrinsic radiation damage}

Models considering the displacement of carbon atoms by high energy electrons have been proposed, \cite{Campbell2000RadiationIrradiation, Losero2023CreationIrradiation} however the vacancy profile predicted for 200~keV electron irradiation (fig. 1(d) of Losero et al. \cite{Losero2023CreationIrradiation}), does not match that measured in our (see fig. \ref{Vdepth}) or other experimental results \cite{RuiangGuoKaiyueWang2020OpticalDiamond}. For 200~keV incident electrons, the probability that an ejected carbon atom has sufficient energy to displace further carbon atoms is negligible and hence secondary cascade processes for vacancy and interstitial production can be ignored. During room temperature electron irradiation interstitial complexes are produced, \cite{Baker1997ElectronIn, Hunt2000IdentificationDiamond, Hunt2000EPRDiamond, Goss2001Self-interstitialDiamond} indicating that at least one form of self-interstitial is mobile \cite{Newton2002Recombination-enhancedDiamond}. 
The annealing of interstitial atoms under conditions during electron irradiation is different to the thermally activated annealing of the $\langle 001 \rangle$-split self-interstitial (C\textsubscript{i$\langle 001 \rangle$}); under electron irradiation a highly mobile interstitial is produced by charge transfer or electronic excitation \cite{Newton2002Recombination-enhancedDiamond}. 

The data in fig. \ref{Vdepth} shows that the maximum depth at which significant concentrations of vacancies are detected is consistent with energy loss due to electronic excitation; after propagating approximately 38~\unit{\um} in diamond \cite{erasmus2014simulation} the energy of a 200~keV incident electron has decreased below the displacement threshold.
The estimated 200~keV electron dose was $10^{19}$~\ecmtwo\ and comparison of the vacancy distribution profile determined from photoluminescence and bulk optical absorption measurements enabled the vacancy production to be determined (section III.A). At the surface the maximum vacancy concentration was $(1.1 \pm 0.2)\times 10^{17}$~V/cm$^3$ equating to a production rate of 0.01~V/e$^-$~cm. 
This is 50 times less than recent modelling predicts \cite{Smith2004StructureDiamond}. Thus either the experimental electron dose has been overestimated, or the modelling of vacancy production is incorrect. Even if the quantitative discrepancy in vacancy production could be in part explained by overestimation of the electron dose, theory does not produce the observed vacancy production profile. 
To bring theory and experiment into agreement for vacancy production by low energy electron irradiation it is suggested that the modelling must include the contribution of electronic excitation to interstitial migration, \cite{Newton2002Recombination-enhancedDiamond} and related changes to the barrier between vacancy-interstitial recombination must be considered to account for the measured profile.

Thermal annealing at 400 and 600\textcelsius\ for one hour did not change the concentration of vacancies produced by 200~keV electron irradiation or the profile (fig. \ref{VdepthAnneal}), as measured by the GR1 PL ZPL and sideband. The former is surprising since absorption studies on diamond samples irradiated with 1-2~MeV electrons at room temperature and below, show that when the \Cione\ is annealed out there is a corresponding reduction in vacancies \cite{Kiflawi2007ElectronDiamond}.
The same study showed absorption using the area under the GR1 ZPL underestimated the \vn\ concentration by approximately 25\% due to the presence of strained vacancies (i.e. those with a nearby interstial). The study showed once these close V-\Cione\ pairs annealed out such underestimation was no longer a problem.

We have not been able to measure the \Cione\ concentration in this sample, but the lack of change of the vacancy PL after annealing at 600\textcelsius\ suggests that either the \Cione\ concentration is much lower than that of vacancies, or the vacancies close to interstitials (presumably the first to anneal out) do not luminesce. 
Energy transfer between close V-\Cione\ pairs is entirely possible, and \Cione\ does not luminescence, so such energy transfer is followed by non radiative de-excitation \cite{Smith2004StructureDiamond}. 
Alternatively, the lower (than predicted by theory) vacancy production measured may be accounted for by higher than expected vacancy-interstitial recombination, with fewer close vacancy interstitial pairs surviving, during low energy electron irradiation. If this is accompanied by higher production rates of stable interstitial aggregates then indeed the production of isolated interstitials may be less, for low energy electron irradiation. This warrants further study to see if the production efficiency of isolated interstitials relative to vacancies varies with electron irradiation energy or vacancies in close  V-\Cione\ pairs do not luminesce.

\subsection{Vacancy clusters}
At 800\textcelsius, in the absence of impurities, vacancies are lost to multi-vacancy complexes, also termed vacancy clusters.   
The structure, electron and optical properties of multi-vacancy point defects in diamond have been considered both experimentally and theoretically. 
Early experimental work was interpreted as the existence of a family of $\{110\}$ oriented vacancy chains, \cite{Lomer1973ElectronTemperatures} but more recent theoretical studies indicated that $\{110\}$ planar vacancy chains larger than three-vacancies are not energetically the most stable structures. Exceptionally stable clusters are composed of 6 (atoms in a chair configuration are removed to form an ideal hexa-vacancy cluster), 10 and 14 vacancies are predicted, but this stability was not attributed to re-bonding as was the case in silicon \cite{Hounsome2005OpticalDiamond}. 
The $\{111\}~\pi $-bonded vacancy disc (a double $\{111\}$ plane of carbon atoms removed) has a much lower in formation energy per vacancy than any of the multi-vacancy point defects \cite{Hounsome2005OpticalDiamond, Jones2007ElectricalDiamond}.
Thus if a large quantity of vacancies is introduced by irradiation damage, then upon annealing at temperatures where vacancies are mobile, but the clusters stable, we would expect the formation of a variety of vacancy clusters. The smaller clusters (e.g. \Vtwo\ and \Vthree ) will in turn be lost as they trap mobile vacancies to make larger clusters and ultimately vacancy discs.

In silicon for the [111] di-vacancy to reorient the two vacancies have to separate by one lattice spacing and then reunite but with a differently oriented axis \cite{Corbett1965ProductionSilicon}. The fact that this occurs with no loss of di-vacancies shows that there is a substantial binding energy between the two vacancies and demonstrates that in silicon the di-vacancy can reorientate and diffuse without permanently dissociating. 
Such a reorientation of di-vacancies in diamond has not, to the best of our knowledge, been measured. The neutral nearest neighbour di-vacancy (\Vtwo ) in diamond, \cite{Twitchen1999CorrelationEPR} anneals in as the isolated vacancy anneals out, but does not persist as the temperature is increased and itself anneals out rapidly at temperatures above about 900\textcelsius\ \cite{Baker1999CentresDiamond}. 
The 733~nm centre also anneals in as the isolated vacancy anneals out and anneals out at temperatures above about 1200\textcelsius . This defect has been previously attributed to a di-vacancy defect in diamond, \cite{Steeds2014AnnealingSamples} but the structure has not been determined.

\subsection{Nitrogen-vacancy (NV) and nitrogen-di-vacancy (\nvv ) defects}
If present, it is well known that the substitutional nitrogen defect is an effective trap for the vacancy, producing the NV defect. Above 1400\textcelsius\ the NV defects readily migrate and can be trapped by remaining substitutional nitrogen defect to produce \nnv . 
The energy required to completely break a NV defect (i.e., for the vacancy to escape Coulombic attraction and the nitrogen-induced strain field) is greater than that required for a sequence of local vacancy emission and re-trapping steps \cite{Ashfold2020NitrogenDiamond}.
In this process the vacancy interchanges place with a neighbouring carbon, and then either returns to re-form the original NV or moves further away into the third nearest neighbour site of the nitrogen. From this position it does not escape but returns to the nitrogen; however, in doing so could take an alternate route and produce a differently oriented NV defect. 
NV reorientation has been observed at temperatures as low as 1050\textcelsius , \cite{Chakravarthi2020WindowCenters} but for NV diffusion the nitrogen and vacancy must additionally swap places. Significant numbers of NV defects survive annealing at 1200\textcelsius\ and this annealing temperature is often used in an effort to remove di-vacancy defects and optimise NV production. 

In recent work using ion implantation the relative production efficiencies of NV and \Vtwo\ defects have been investigated \cite{Santonocito2024NVModelling}. Additionally, when the concentration of vacancies is high it is necessary to consider the production of \nvv\ since NV is also a trap for vacancies. The most likely structure of \nvv\ is a di-vacancy with one of the six carbon neighbours replaced with a nitrogen \cite{Kuganathan2023VacancyDiamond}. The consequence of this is that NV production is reduced in the presence of excess vacancies.

\section{\label{sec:Modell}Modelling of vacancy annealing}
\begin{figure}
\includegraphics[width=\columnwidth]{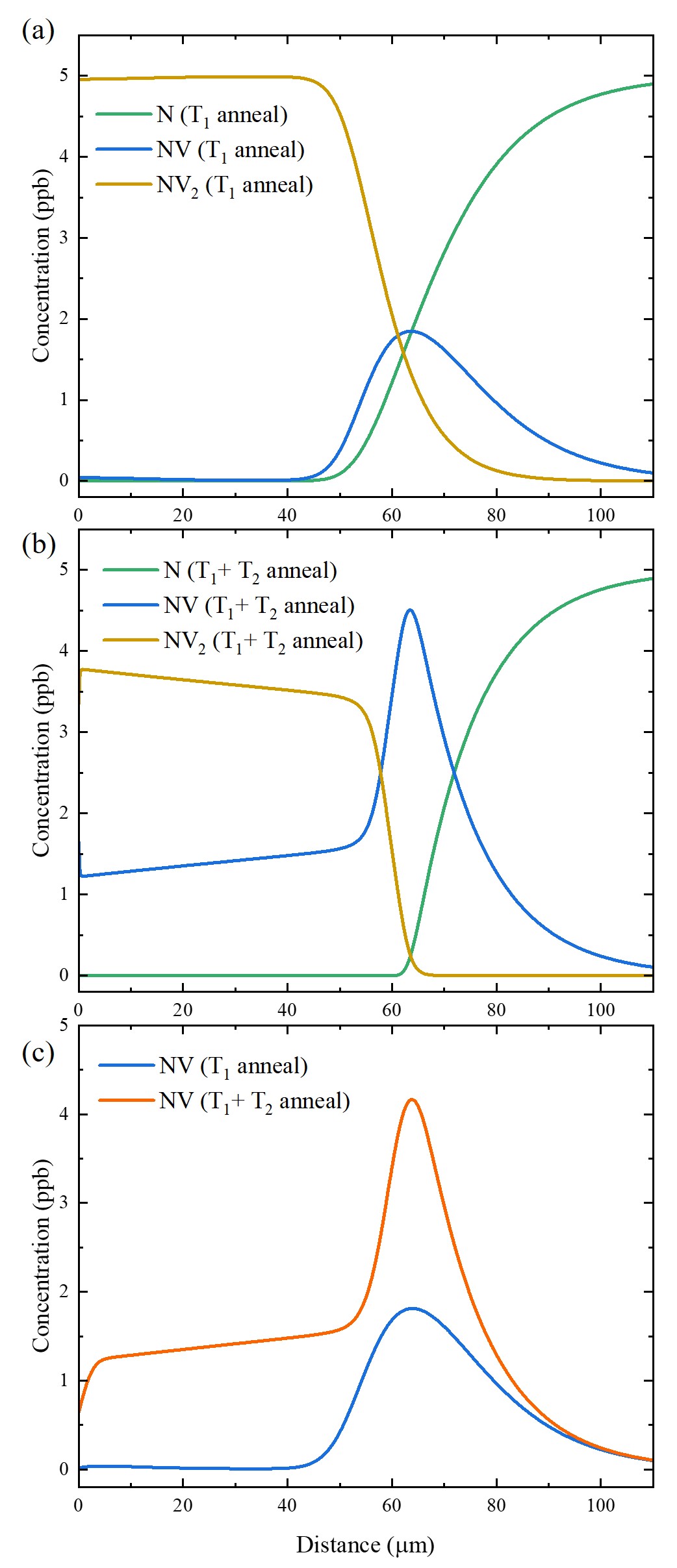}
\caption{\label{SimAnnealNV} (a) Simulation of the N, NV and \nvv\ concentrations after 1 hour annealing at $T_1$=1000\textcelsius . (b) Simulation of the N, NV and \nvv\ concentrations after 1 hour annealing at $T_1$=1000\textcelsius\ followed by 1 hour annealing at $T_2$=1200\textcelsius . (c) NV concentrations from (a \& b) convolved with a Gaussian to approximate the spectrometer response function.}
\end{figure}

\subsection{Chemical kinetics}
In order to explain/investigate the vacancy aggregation pathways a simple chemical kinetics model of the vacancy annealing was implemented. In the simple model described here, it is assumed that small vacancy clusters (\Vtwo, \Vthree, \Vfour) anneal in by vacancy aggregation, but are not themselves mobile and \Vtwo\ and \Vthree\ are lost at higher temperatures by dissociation. \Vfour\ is assumed to be stable at the annealing temperatures studied. The simulation could be extended to include diffusion of the multi-vacancy aggregates, and formation of larger clusters of vacancies, but there is insufficient data against which to constrain or confirm the former, and the latter would not change our interpretation of the data for isolated vacancies and the nitrogen-vacancy defect. Additionally, we assume that \nvv\ dissociates into a NV and a V at an annealing temperature below the annealing temperature at which the NV defect is lost.

The model consists of a set of simultaneous equations describing defect migration and complex formation (aggregation). For simplicity, we start with isolated vacancies (V) and single-substitutional nitrogen (\Ns )  defects, and do not consider different charge states of any species. 
Vacancies and \Ns\ are assumed not to interact with interstitials or interstitial complexes, but diffuse at sufficiently high annealing temperatures. The vacancies aggregate to form multi-vacancy complexes (\Vtwo , \Vthree , and \Vfour ) or can aggregate with \Ns\ to form nitrogen-vacancy (NV) and nitrogen-di-vacancy (\nvv ) defects and all multi-vacancy complexes are assumed to be immobile. 
Two temperature regimes are considered: $T_1$ and $T_2$, where $T_2 > T_1$. At annealing temperature $T_1$ it is assumed that \Vtwo , \Vthree , \Vfour , NV and \nvv\ are stable and do not dissociate. At annealing temperature $T_2$ it is assumed that \Vtwo , \Vthree\ and \nvv\ dissociate (e.g. \Vtwo $\rightarrow$~V+V, \Vthree $\rightarrow$~\Vtwo\ +V and \nvv$\rightarrow$~NV+V) but NV and \Vfour\ are both stable.
The boundary conditions are assumed to be infinite sinks for vacancies and the finite difference method is used to calculate concentrations of all defects during annealing. The full set of rate equations used to implement the model can be found in the supporting information, but the simultaneous partial differential equations are as follows: 
\begin{align}
\begin{split}
    \frac{\partial [\text{V}] }{\partial t}=D_V \frac{\partial ^2 [\text{V}] }{\partial x^2}-k_2 [\text{V}]^2 - k_3 [\text{V}][\text{\Vtwo}] - k_4 \;[\text{V}][\text{\Vthree}] \\
     +2 l_2 [\text{\Vtwo}] +l_3[\text{\Vthree}] - k_{NV}[\text{V}][\text{\Ns}]\\
     - k_{NV2}[\text{V}][\text{NV}] + l_{NV2}[\text{\nvv}]
    \label{eq: dV}\\
\end{split}
\end{align}
\begin{align}
    \frac{\partial [\text{\Vtwo}] }{\partial t}&= + \frac{k_2}{2} [\text{V}]^2 -  k_3 [\text{V}][\text{\Vtwo}] -l_2[\text{\Vtwo}] +l_3[\text{\Vthree}]
    \label{eq: dV2}\\
    \frac{\partial [\text{\Vthree}] }{\partial t}&= + k_3 [\text{V}]^2 -k_4[\text{V}][\text{\Vthree}]  -l_3[\text{\Vthree}]\\
    \frac{\partial [\text{\Vfour}] }{\partial t}&= +k_4[\text{V}][\text{\Vthree}] \\
    \frac{\partial [\text{\Ns}] }{\partial t}&= -k_{NV}[\text{V}][\text{\Ns}]\\
    \frac{\partial [\text{NV}] }{\partial t}&= + k_{NV}[\text{V}][\text{\Ns}] - k_{NV2}[\text{V}][\text{NV}] + l_{NV2}[\text{\nvv}]\\
    \frac{\partial [\text{\nvv}] }{\partial t}&= + k_{NV2}[\text{V}][\text{NV}] - l_{NV2}[\text{\nvv}]
    \label{eq: dNV2}
\end{align}
where [S] is used to denote the concentration of species S, and $t$ and $x$ are time and distance respectively. $D_V$ is the diffusion coefficient of vacancies and $k_s$ and $l_s$ are the rate constants associated with the various aggregation and dissociation processes involved. 
The vacancy diffusion coefficient and all the rate constants are assumed to have an Arrhenius temperature dependence, with an activation energy appropriate for the specific reaction/process.

\subsection{Simulation results}
To demonstrate the output of this model at a temperature $T_1=1000$~\textcelsius\ we choose $k_2 =k_3 =k_4= 2\times10^{-5}$~s$^{-1}$/ppb, $k_{NV}=k_{NV2}=2\times 10^{-4}$~s$^{-1}$/ppb, $l_2=l_3=2\times 10^{-6}$~s$^{-1}$, $l_{NV2}=3\times 10^{-6}$~s$^{-1}$, and $D_V=1.5\times 10^{-6}$~\unit{\um}$^2$s$^{-1}$ \cite{Onoda2017DiffusionDiamond}.
For an initial vacancy concentration distribution decaying exponentially into the sample ([V]=V\textsubscript{0}exp($-x/x_0$), where V\textsubscript{0}=1000~ppb and $x_0$=12\unit{\um}), and a uniform doping with single substitutional nitrogen [\Ns]=5~ppb, 
we find after a 1 hour anneal there is a small peak in the concentration of vacancies approximately 40~\unit{\um} into the diamond, displayed in fig. \ref{VdepthAnneal}(d). 
The position and intensity of the vacancy peak can be changed by adjusting the rate constants, but we have resisted the temptation to fit the experimental data as there are too many variable parameters and too little data to constrain the fit. Further experimental studies with different starting concentrations of both V and \Ns , and a range of annealing times and temperatures are required to constrain the fitting parameters. 
However, what we find is that the peak in the concentration of V cannot be produced by simple second order kinetics for the loss of vacancies  (V+V$\rightarrow$~\Vtwo). It is the production of multiple vacancy complexes, fig \ref{VdepthAnneal}(d), (e.g. traps for vacancies) that results in a peak in vacancy concentration significantly far into the diamond.  

The resulting profiles of [\Ns], [NV], and [\nvv] after the simulated anneal are shown in fig. \ref{SimAnnealNV}(a).
For the starting V concentration we can see that at depths $<63.5$~\unit{\um} [V]$>$\Ns] and at depths $>63.5$~\unit{\um} [\Ns]$>$[V]. 
Indeed at the surface [V]=200$\times$[\Ns] ([V]$\gg$[\Ns]) and at a depth of 100~\unit{\um} [\Ns]$\approx 20\times$[V] ([\Ns]$>$[V]). 
Setting $k_{NV},~k_{NV2}>k_2,~k_3,~k_4$ ensures that although even though [V]$\gg$[\Ns] in the first $\approx40$~\unit{\um} the vacancies are efficiently trapped by the nitrogen, and nearly all the \Ns\ is converted to \nvv. 
At greater depths, there are insufficient vacancies to achieve this conversion and measurable concentrations of NV are produced. Moving even further into the diamond the concentration of NV reaches a maximum at $\approx$64~\unit{\um} into the diamond and then decreases, as there are now insufficient vacancies to produce significant concentrations of NV. Deep in the diamond all the nitrogen remains as \Ns . 
Again the size and position of the peak in NV concentration can be altered by changing the rate constants, but for the same reasons as discussed above no fit to the experimental data has been attempted. Nevertheless, this simple approach explains the experimentally observed NV PL profile (fig. \ref{fig:NVanneal}). 

To simulate a second anneal at $T_2$=1200\textcelsius\ the rate and diffusion constants must be recalculated to account for the increase in temperature.
For the purpose of simulation we have assumed an activation energy for the formation of \Vtwo, \Vthree, \Vfour, NV, and \nvv\ of 2.2~eV, \cite{Davies1992Vacancy-relatedDiamond} an activation energy for vacancy diffusion of 2.0~eV, a \Vtwo\ and \Vthree\ dissociation energy of 3.5~eV, \cite{Slepetz2014DivacanciesMechanism} and a \nvv\ dissociation energy of 3.8~eV. 
Thus on increasing the temperature from 1000\textcelsius\ to 1200\textcelsius\ the rate constants become $k_2=k_3=k_4= 1.4\times10^{-3}$~s$^{-1}$/ppb, $k_{NV}=k_{NV2}=1.4\times 10^{-2}$~s$^{-1}$/ppb, $l_2=l_3=1.7\times 10^{-3}$~s$^{-1}$, l$l_{NV2}=4.6\times 10^{-3}$~s$^{-1}$ and $D_V=7.2\times 10^{-5}$~\unit{\um}$^2$s$^{-1}$. 
If the 1 hour anneal at $T_1=1000$\textcelsius\ is followed by a further 1 hour at $T_2=1200$\textcelsius, then the simulation shows that the 97\% of vacancies are now in the form of \Vfour\ clusters, with only a very few V, \Vtwo , and \Vthree\ remaining. If \Vthree\ is responsible for the 733~nm emission (fig. \ref{733depthAnneal}) then then the loss is overestimated in this simulation. Fig. \ref{SimAnnealNV}(b) shows the predicted concentration of \Ns, NV, and \nvv\ after this second anneal. 
The dissociation of \nvv\ results in an increase in the concentration of NV, especially in the first 40~\unit{\um} of the diamond (fig. \ref{SimAnnealNV}(b) and this seen in the experimental PL intensity data (fig. \ref{fig:NVanneal}(a) and \ref{fig:NVanneal}(c)). 

\subsection{Charge transfer}
NV centres give insight into the surrounding donor concentrations by their charge state. If there are nearby donors it is energetically favourable for NV centres to be negatively charged; as stated \nvn $\rightarrow$  \nvm\ can also be achieved via photoionisation. The differences in the \nvn\ and \nvm\ depth profiles with different excitation in fig. \ref{fig:NVanneal} are a result of such charge transfer. 

In the experimental data the emission from \nvn\ dominates over that from \nvm\ in the first 40\unit{\um} (fig. \ref{fig:NVanneal}(a) and  \ref{fig:NVanneal}(c)) which is consistent with the simulated higher concentration of competing \nvv\ traps. 
Overall the relative emission from \nvm\ to that from \nvn\ has increased on annealing at 1200\textcelsius\ suggesting the number of competing electron traps has been reduced. Photoionization of \nvm\ with 488~nm excitation is again apparent (fig. \ref{fig:NVanneal}(a) and  \ref{fig:NVanneal}(c)). Fig. \ref{SimAnnealNV}(c) shows the simulated total NV concentration after the first anneal at $T_1$, and after both anneals at $T_1$ and  $T_2$. The concentration profiles have been convolved with a Gaussian (FWHH 5~\unit{\um}) to approximate the spectrometer spatial response function. 
No quantitative agreement between fig. \ref{SimAnnealNV}(c) and fig. \ref{fig:NVanneal} is claimed, but qualitatively the simulation is predicting the observed behaviour. There is no merit to simply varying the rate constants to improve this agreement, without more data to constrain these parameters. 

It can be seen the optimal \vn\ concentration for \nvm\ creation efficiency in the sample is the concentration present at a depth of 60$\pm 5$~\unit{\um}, i.e. $5\pm 3$~ppb \vn .

\section{Conclusions and future work}
The radiation damage depth profile for low energy 200~keV electrons in electronic grade diamond has been measured and the creation efficiency of the vacancy production has been quantified. 
The depth profile was found to follow an exponential decay, with a decay constant of $12\pm1~$\unit{\um}. This is not what has been predicted by theory \cite{Campbell2000RadiationIrradiation, Losero2023CreationIrradiation} and more theoretical input is required to reconcile this. 
A dose of 10$^{19}~$\ecmtwo\ resulted in a peak vacancy concentration of $(1.1\pm 0.2) \times 10^{17}$~V/cm$^3$ at the surface. 

Assuming vacancies anneal to form vacancy clusters results in a change in the vacancy distribution consisting of a small surviving peak in the concentration of \vn below the surface. The position and size of the peak depend on the rate constants used in the model. In this work no attempt has been made to fit the experimental measurements, as more data is required.

The simulation supports assignments of the 733~nm ZPL PL feature to a vacancy cluster, as the concentration depth profile follows that of a vacancy cluster (see supporting information) and we speculate that this might be \Vthree. 

Our simple model predicts the observed NV distributions, but relies on the formation of \nvv . At higher temperatures \nvv\ dissociates, additionally changing the NV distribution. 
\nvv\ plays an important role in determining the charge state of the NV defect by acting as an electron trap.

Near the surface ($<$20~\unit{\um}) there are many more vacancies than substitutional nitrogen; experimental data suggests 200 to 1. In this regime, we do not successfully form a high concentration of NV centres and instead form \nvv . This could explain the relatively poor yield of NV defects by ion implantation; it is well known that correct surface termination is necessary for maximising yield however implantation also results in a regime where [V]$\gg$[\Ns].
For example, implantation of diamond with $2\times 10^9$ 30~keV nitrogen ions/cm$^2$  results in approximately 5~ppb ($9\times10^{14}$~cm$^{-3}$) of nitrogen atoms (calculated using SRIM) at a depth of 40~nm, whereas the concentration of vacancies produced at this depth is approximately 400~ppb. Thus [V]$\gg$[\Ns] and our analysis suggests that the production efficiency of NV on annealing would be significantly reduced by the formation of \nvv . Santonocito et al. \cite{Santonocito2024NVModelling} also cite the production of \nvv\ during ion implantation and annealing. Further study and detection of \nvv\ (in either the neutral or negative charge state) must be a priority in order to understand how NV production may be optimised.

In electronic grade diamond with a concentration of 5~ppb \Ns\ the optimal \vn\ concentration for \nvm\ creation is 5$\pm $3~ppb. To optimise NV concentration close to the surface after an electron irradiation dose of $10^{19}$~\ecmtwo\, removal of the top 30-50~\unit{\um} would be recommended in such samples. 

Further work to study both the electron dose and energy dependence of the depth profiles for vacancy creation and NV production would be beneficial, and variation of the nitrogen concentration in the sample is required. 

\begin{acknowledgments}
We wish to acknowledge the support of the spectroscopy RTP at the University of Warwick.
We gratefully acknowledge the support of the DOE Q-NEXT Center (Grant No. DOE 1F-60579) and the use of shared facilities of the UCSB Quantum Foundry (NSF DMR1906325). A.B.J. acknowledges support  from NSF QLCI program through Grant No. OMA-2016245.
\end{acknowledgments}


\bibliography{Nreferences, extra}

\begin{thebibliography}{52}%
\makeatletter
\providecommand \@ifxundefined [1]{%
 \@ifx{#1\undefined}
}%
\providecommand \@ifnum [1]{%
 \ifnum #1\expandafter \@firstoftwo
 \else \expandafter \@secondoftwo
 \fi
}%
\providecommand \@ifx [1]{%
 \ifx #1\expandafter \@firstoftwo
 \else \expandafter \@secondoftwo
 \fi
}%
\providecommand \natexlab [1]{#1}%
\providecommand \enquote  [1]{``#1''}%
\providecommand \bibnamefont  [1]{#1}%
\providecommand \bibfnamefont [1]{#1}%
\providecommand \citenamefont [1]{#1}%
\providecommand \href@noop [0]{\@secondoftwo}%
\providecommand \href [0]{\begingroup \@sanitize@url \@href}%
\providecommand \@href[1]{\@@startlink{#1}\@@href}%
\providecommand \@@href[1]{\endgroup#1\@@endlink}%
\providecommand \@sanitize@url [0]{\catcode `\\12\catcode `\$12\catcode `\&12\catcode `\#12\catcode `\^12\catcode `\_12\catcode `\%12\relax}%
\providecommand \@@startlink[1]{}%
\providecommand \@@endlink[0]{}%
\providecommand \url  [0]{\begingroup\@sanitize@url \@url }%
\providecommand \@url [1]{\endgroup\@href {#1}{\urlprefix }}%
\providecommand \urlprefix  [0]{URL }%
\providecommand \Eprint [0]{\href }%
\providecommand \doibase [0]{https://doi.org/}%
\providecommand \selectlanguage [0]{\@gobble}%
\providecommand \bibinfo  [0]{\@secondoftwo}%
\providecommand \bibfield  [0]{\@secondoftwo}%
\providecommand \translation [1]{[#1]}%
\providecommand \BibitemOpen [0]{}%
\providecommand \bibitemStop [0]{}%
\providecommand \bibitemNoStop [0]{.\EOS\space}%
\providecommand \EOS [0]{\spacefactor3000\relax}%
\providecommand \BibitemShut  [1]{\csname bibitem#1\endcsname}%
\let\auto@bib@innerbib\@empty
\bibitem [{\citenamefont {Doherty}\ \emph {et~al.}(2013)\citenamefont {Doherty}, \citenamefont {Manson}, \citenamefont {Delaney}, \citenamefont {Jelezko}, \citenamefont {Wrachtrup},\ and\ \citenamefont {Hollenberg}}]{Doherty2013TheDiamond}%
  \BibitemOpen
  \bibfield  {author} {\bibinfo {author} {\bibfnamefont {M.~W.}\ \bibnamefont {Doherty}}, \bibinfo {author} {\bibfnamefont {N.~B.}\ \bibnamefont {Manson}}, \bibinfo {author} {\bibfnamefont {P.}~\bibnamefont {Delaney}}, \bibinfo {author} {\bibfnamefont {F.}~\bibnamefont {Jelezko}}, \bibinfo {author} {\bibfnamefont {J.}~\bibnamefont {Wrachtrup}},\ and\ \bibinfo {author} {\bibfnamefont {L.~C.}\ \bibnamefont {Hollenberg}},\ }\bibfield  {title} {\bibinfo {title} {{The nitrogen-vacancy colour centre in diamond}},\ }\href {https://doi.org/10.1016/j.physrep.2013.02.001} {\bibfield  {journal} {\bibinfo  {journal} {Physics Reports}\ }\textbf {\bibinfo {volume} {528}},\ \bibinfo {pages} {1} (\bibinfo {year} {2013})}\BibitemShut {NoStop}%
\bibitem [{\citenamefont {Wolf}\ \emph {et~al.}(2015)\citenamefont {Wolf}, \citenamefont {Neumann}, \citenamefont {Nakamura}, \citenamefont {Sumiya}, \citenamefont {Ohshima}, \citenamefont {Isoya},\ and\ \citenamefont {Wrachtrup}}]{Wolf2015SubpicoteslaMagnetometry}%
  \BibitemOpen
  \bibfield  {author} {\bibinfo {author} {\bibfnamefont {T.}~\bibnamefont {Wolf}}, \bibinfo {author} {\bibfnamefont {P.}~\bibnamefont {Neumann}}, \bibinfo {author} {\bibfnamefont {K.}~\bibnamefont {Nakamura}}, \bibinfo {author} {\bibfnamefont {H.}~\bibnamefont {Sumiya}}, \bibinfo {author} {\bibfnamefont {T.}~\bibnamefont {Ohshima}}, \bibinfo {author} {\bibfnamefont {J.}~\bibnamefont {Isoya}},\ and\ \bibinfo {author} {\bibfnamefont {J.}~\bibnamefont {Wrachtrup}},\ }\bibfield  {title} {\bibinfo {title} {{Subpicotesla Diamond Magnetometry}},\ }\href {https://doi.org/10.1103/PhysRevX.5.041001} {\bibfield  {journal} {\bibinfo  {journal} {Physical Review X}\ }\textbf {\bibinfo {volume} {5}},\ \bibinfo {pages} {041001} (\bibinfo {year} {2015})}\BibitemShut {NoStop}%
\bibitem [{\citenamefont {Dolde}\ \emph {et~al.}(2011)\citenamefont {Dolde}, \citenamefont {Fedder}, \citenamefont {Doherty}, \citenamefont {N{\"{o}}bauer}, \citenamefont {Rempp}, \citenamefont {Balasubramanian}, \citenamefont {Wolf}, \citenamefont {Reinhard}, \citenamefont {Hollenberg}, \citenamefont {Jelezko},\ and\ \citenamefont {Wrachtrup}}]{Dolde2011Electric-fieldSpins}%
  \BibitemOpen
  \bibfield  {author} {\bibinfo {author} {\bibfnamefont {F.}~\bibnamefont {Dolde}}, \bibinfo {author} {\bibfnamefont {H.}~\bibnamefont {Fedder}}, \bibinfo {author} {\bibfnamefont {M.~W.}\ \bibnamefont {Doherty}}, \bibinfo {author} {\bibfnamefont {T.}~\bibnamefont {N{\"{o}}bauer}}, \bibinfo {author} {\bibfnamefont {F.}~\bibnamefont {Rempp}}, \bibinfo {author} {\bibfnamefont {G.}~\bibnamefont {Balasubramanian}}, \bibinfo {author} {\bibfnamefont {T.}~\bibnamefont {Wolf}}, \bibinfo {author} {\bibfnamefont {F.}~\bibnamefont {Reinhard}}, \bibinfo {author} {\bibfnamefont {L.~C.}\ \bibnamefont {Hollenberg}}, \bibinfo {author} {\bibfnamefont {F.}~\bibnamefont {Jelezko}},\ and\ \bibinfo {author} {\bibfnamefont {J.}~\bibnamefont {Wrachtrup}},\ }\bibfield  {title} {\bibinfo {title} {{Electric-field sensing using single diamond spins}},\ }\href {https://doi.org/10.1038/nphys1969} {\bibfield  {journal} {\bibinfo  {journal} {Nature Physics}\ }\textbf {\bibinfo {volume} {7}},\ \bibinfo {pages} {459} (\bibinfo {year}
  {2011})}\BibitemShut {NoStop}%
\bibitem [{\citenamefont {Schirhagl}\ \emph {et~al.}(2014)\citenamefont {Schirhagl}, \citenamefont {Chang}, \citenamefont {Loretz},\ and\ \citenamefont {Degen}}]{Schirhagl2014Nitrogen-vacancyBiology}%
  \BibitemOpen
  \bibfield  {author} {\bibinfo {author} {\bibfnamefont {R.}~\bibnamefont {Schirhagl}}, \bibinfo {author} {\bibfnamefont {K.}~\bibnamefont {Chang}}, \bibinfo {author} {\bibfnamefont {M.}~\bibnamefont {Loretz}},\ and\ \bibinfo {author} {\bibfnamefont {C.~L.}\ \bibnamefont {Degen}},\ }\bibfield  {title} {\bibinfo {title} {{Nitrogen-vacancy centers in diamond: Nanoscale sensors for physics and biology}},\ }\href {https://doi.org/10.1146/annurev-physchem-040513-103659} {\bibfield  {journal} {\bibinfo  {journal} {Annual Review of Physical Chemistry}\ }\textbf {\bibinfo {volume} {65}},\ \bibinfo {pages} {83} (\bibinfo {year} {2014})}\BibitemShut {NoStop}%
\bibitem [{\citenamefont {Kucsko}\ \emph {et~al.}(2013)\citenamefont {Kucsko}, \citenamefont {Maurer}, \citenamefont {Yao}, \citenamefont {Kubo}, \citenamefont {Noh}, \citenamefont {Lo}, \citenamefont {Park},\ and\ \citenamefont {Lukin}}]{Kucsko2013Nanometre-scaleCell}%
  \BibitemOpen
  \bibfield  {author} {\bibinfo {author} {\bibfnamefont {G.}~\bibnamefont {Kucsko}}, \bibinfo {author} {\bibfnamefont {P.~C.}\ \bibnamefont {Maurer}}, \bibinfo {author} {\bibfnamefont {N.~Y.}\ \bibnamefont {Yao}}, \bibinfo {author} {\bibfnamefont {M.}~\bibnamefont {Kubo}}, \bibinfo {author} {\bibfnamefont {H.~J.}\ \bibnamefont {Noh}}, \bibinfo {author} {\bibfnamefont {P.~K.}\ \bibnamefont {Lo}}, \bibinfo {author} {\bibfnamefont {H.}~\bibnamefont {Park}},\ and\ \bibinfo {author} {\bibfnamefont {M.~D.}\ \bibnamefont {Lukin}},\ }\bibfield  {title} {\bibinfo {title} {{Nanometre-scale thermometry in a living cell}},\ }\href {https://doi.org/10.1038/nature12373} {\bibfield  {journal} {\bibinfo  {journal} {Nature}\ }\textbf {\bibinfo {volume} {500}},\ \bibinfo {pages} {54} (\bibinfo {year} {2013})}\BibitemShut {NoStop}%
\bibitem [{\citenamefont {Riedel}\ \emph {et~al.}(2017)\citenamefont {Riedel}, \citenamefont {S{\"{o}}llner}, \citenamefont {Shields}, \citenamefont {Starosielec}, \citenamefont {Appel}, \citenamefont {Neu}, \citenamefont {Maletinsky},\ and\ \citenamefont {Warburton}}]{Riedel2017DeterministicDiamond}%
  \BibitemOpen
  \bibfield  {author} {\bibinfo {author} {\bibfnamefont {D.}~\bibnamefont {Riedel}}, \bibinfo {author} {\bibfnamefont {I.}~\bibnamefont {S{\"{o}}llner}}, \bibinfo {author} {\bibfnamefont {B.~J.}\ \bibnamefont {Shields}}, \bibinfo {author} {\bibfnamefont {S.}~\bibnamefont {Starosielec}}, \bibinfo {author} {\bibfnamefont {P.}~\bibnamefont {Appel}}, \bibinfo {author} {\bibfnamefont {E.}~\bibnamefont {Neu}}, \bibinfo {author} {\bibfnamefont {P.}~\bibnamefont {Maletinsky}},\ and\ \bibinfo {author} {\bibfnamefont {R.~J.}\ \bibnamefont {Warburton}},\ }\bibfield  {title} {\bibinfo {title} {{Deterministic enhancement of coherent photon generation from a nitrogen-vacancy center in ultrapure diamond}},\ }\href {https://doi.org/10.1103/PhysRevX.7.031040} {\bibfield  {journal} {\bibinfo  {journal} {Physical Review X}\ }\textbf {\bibinfo {volume} {7}},\ \bibinfo {pages} {031040} (\bibinfo {year} {2017})}\BibitemShut {NoStop}%
\bibitem [{\citenamefont {Aharonovich}\ and\ \citenamefont {Neu}(2014)}]{Aharonovich2014DiamondNanophotonics}%
  \BibitemOpen
  \bibfield  {author} {\bibinfo {author} {\bibfnamefont {I.}~\bibnamefont {Aharonovich}}\ and\ \bibinfo {author} {\bibfnamefont {E.}~\bibnamefont {Neu}},\ }\bibfield  {title} {\bibinfo {title} {{Diamond nanophotonics}},\ }\href {https://doi.org/10.1002/adom.201400189} {\bibfield  {journal} {\bibinfo  {journal} {Advanced Optical Materials}\ }\textbf {\bibinfo {volume} {2}},\ \bibinfo {pages} {911} (\bibinfo {year} {2014})}\BibitemShut {NoStop}%
\bibitem [{\citenamefont {Poem}\ \emph {et~al.}(2015)\citenamefont {Poem}, \citenamefont {Weinzetl}, \citenamefont {Klatzow}, \citenamefont {Kaczmarek}, \citenamefont {Munns}, \citenamefont {Champion}, \citenamefont {Saunders}, \citenamefont {Nunn},\ and\ \citenamefont {Walmsley}}]{Poem2015BroadbandDiamond}%
  \BibitemOpen
  \bibfield  {author} {\bibinfo {author} {\bibfnamefont {E.}~\bibnamefont {Poem}}, \bibinfo {author} {\bibfnamefont {C.}~\bibnamefont {Weinzetl}}, \bibinfo {author} {\bibfnamefont {J.}~\bibnamefont {Klatzow}}, \bibinfo {author} {\bibfnamefont {K.~T.}\ \bibnamefont {Kaczmarek}}, \bibinfo {author} {\bibfnamefont {J.~H.}\ \bibnamefont {Munns}}, \bibinfo {author} {\bibfnamefont {T.~F.}\ \bibnamefont {Champion}}, \bibinfo {author} {\bibfnamefont {D.~J.}\ \bibnamefont {Saunders}}, \bibinfo {author} {\bibfnamefont {J.}~\bibnamefont {Nunn}},\ and\ \bibinfo {author} {\bibfnamefont {I.~A.}\ \bibnamefont {Walmsley}},\ }\bibfield  {title} {\bibinfo {title} {{Broadband noise-free optical quantum memory with neutral nitrogen-vacancy centers in diamond}},\ }\href {https://doi.org/10.1103/PhysRevB.91.205108} {\bibfield  {journal} {\bibinfo  {journal} {Physical Review B - Condensed Matter and Materials Physics}\ }\textbf {\bibinfo {volume} {91}},\ \bibinfo {pages} {205108} (\bibinfo {year} {2015})}\BibitemShut {NoStop}%
\bibitem [{\citenamefont {Benjamin}\ \emph {et~al.}(2009)\citenamefont {Benjamin}, \citenamefont {Lovett},\ and\ \citenamefont {Smith}}]{Benjamin2009ProspectsSpins}%
  \BibitemOpen
  \bibfield  {author} {\bibinfo {author} {\bibfnamefont {S.}~\bibnamefont {Benjamin}}, \bibinfo {author} {\bibfnamefont {B.}~\bibnamefont {Lovett}},\ and\ \bibinfo {author} {\bibfnamefont {J.}~\bibnamefont {Smith}},\ }\bibfield  {title} {\bibinfo {title} {{Prospects for measurement‐based quantum computing with solid state spins}},\ }\href {https://doi.org/10.1002/lpor.200810051} {\bibfield  {journal} {\bibinfo  {journal} {Laser {\&} Photonics Reviews}\ }\textbf {\bibinfo {volume} {3}},\ \bibinfo {pages} {556} (\bibinfo {year} {2009})}\BibitemShut {NoStop}%
\bibitem [{\citenamefont {Nickerson}\ \emph {et~al.}(2014)\citenamefont {Nickerson}, \citenamefont {Fitzsimons},\ and\ \citenamefont {Benjamin}}]{Nickerson2014FreelyLinks}%
  \BibitemOpen
  \bibfield  {author} {\bibinfo {author} {\bibfnamefont {N.~H.}\ \bibnamefont {Nickerson}}, \bibinfo {author} {\bibfnamefont {J.~F.}\ \bibnamefont {Fitzsimons}},\ and\ \bibinfo {author} {\bibfnamefont {S.~C.}\ \bibnamefont {Benjamin}},\ }\bibfield  {title} {\bibinfo {title} {{Freely Scalable Quantum Technologies Using Cells of 5-to-50 Qubits with Very Lossy and Noisy Photonic Links}},\ }\href {https://doi.org/10.1103/PhysRevX.4.041041} {\bibfield  {journal} {\bibinfo  {journal} {Physical Review X}\ }\textbf {\bibinfo {volume} {4}},\ \bibinfo {pages} {041041} (\bibinfo {year} {2014})}\BibitemShut {NoStop}%
\bibitem [{\citenamefont {Eichhorn}\ \emph {et~al.}(2019)\citenamefont {Eichhorn}, \citenamefont {Mclellan},\ and\ \citenamefont {Bleszynski~Jayich}}]{Eichhorn2019OptimizingSensing}%
  \BibitemOpen
  \bibfield  {author} {\bibinfo {author} {\bibfnamefont {T.~R.}\ \bibnamefont {Eichhorn}}, \bibinfo {author} {\bibfnamefont {C.~A.}\ \bibnamefont {Mclellan}},\ and\ \bibinfo {author} {\bibfnamefont {A.~C.}\ \bibnamefont {Bleszynski~Jayich}},\ }\bibfield  {title} {\bibinfo {title} {{Optimizing the formation of depth-confined nitrogen vacancy center spin ensembles in diamond for quantum sensing}},\ }\href {https://doi.org/10.1103/PhysRevMaterials.3.113802} {\bibfield  {journal} {\bibinfo  {journal} {Physical Review Materials}\ }\textbf {\bibinfo {volume} {3}},\ \bibinfo {pages} {113802} (\bibinfo {year} {2019})}\BibitemShut {NoStop}%
\bibitem [{\citenamefont {Chen}\ \emph {et~al.}(2019)\citenamefont {Chen}, \citenamefont {Griffiths}, \citenamefont {Ishmael}, \citenamefont {Newton}, \citenamefont {Nicley}, \citenamefont {Stephen}, \citenamefont {Lekhai}, \citenamefont {Booth}, \citenamefont {Salter}, \citenamefont {Johnson}, \citenamefont {Green}, \citenamefont {Weng}, \citenamefont {Morley},\ and\ \citenamefont {Smith}}]{Chen2019LaserYield}%
  \BibitemOpen
  \bibfield  {author} {\bibinfo {author} {\bibfnamefont {Y.-C.}\ \bibnamefont {Chen}}, \bibinfo {author} {\bibfnamefont {B.}~\bibnamefont {Griffiths}}, \bibinfo {author} {\bibfnamefont {S.~N.}\ \bibnamefont {Ishmael}}, \bibinfo {author} {\bibfnamefont {M.~E.}\ \bibnamefont {Newton}}, \bibinfo {author} {\bibfnamefont {S.~S.}\ \bibnamefont {Nicley}}, \bibinfo {author} {\bibfnamefont {C.~J.}\ \bibnamefont {Stephen}}, \bibinfo {author} {\bibfnamefont {Y.}~\bibnamefont {Lekhai}}, \bibinfo {author} {\bibfnamefont {M.~J.}\ \bibnamefont {Booth}}, \bibinfo {author} {\bibfnamefont {P.~S.}\ \bibnamefont {Salter}}, \bibinfo {author} {\bibfnamefont {S.}~\bibnamefont {Johnson}}, \bibinfo {author} {\bibfnamefont {B.~L.}\ \bibnamefont {Green}}, \bibinfo {author} {\bibfnamefont {L.}~\bibnamefont {Weng}}, \bibinfo {author} {\bibfnamefont {G.~W.}\ \bibnamefont {Morley}},\ and\ \bibinfo {author} {\bibfnamefont {J.~M.}\ \bibnamefont {Smith}},\ }\bibfield  {title} {\bibinfo {title} {{Laser writing of individual nitrogen-vacancy
  defects in diamond with near-unity yield}},\ }\href {https://doi.org/10.1364/OPTICA.6.000662} {\bibfield  {journal} {\bibinfo  {journal} {Optica}\ }\textbf {\bibinfo {volume} {6}},\ \bibinfo {pages} {662} (\bibinfo {year} {2019})}\BibitemShut {NoStop}%
\bibitem [{\citenamefont {Edmonds}\ \emph {et~al.}(2021)\citenamefont {Edmonds}, \citenamefont {Hart}, \citenamefont {Turner}, \citenamefont {Colard}, \citenamefont {Schloss}, \citenamefont {Olsson}, \citenamefont {Trubko}, \citenamefont {Markham}, \citenamefont {Rathmill}, \citenamefont {Horne-Smith}, \citenamefont {Lew}, \citenamefont {Manickam}, \citenamefont {Bruce}, \citenamefont {Kaup}, \citenamefont {Russo}, \citenamefont {DiMario}, \citenamefont {South}, \citenamefont {Hansen}, \citenamefont {Twitchen},\ and\ \citenamefont {Walsworth}}]{Edmonds2021CharacterisationApplications}%
  \BibitemOpen
  \bibfield  {author} {\bibinfo {author} {\bibfnamefont {A.~M.}\ \bibnamefont {Edmonds}}, \bibinfo {author} {\bibfnamefont {C.~A.}\ \bibnamefont {Hart}}, \bibinfo {author} {\bibfnamefont {M.~J.}\ \bibnamefont {Turner}}, \bibinfo {author} {\bibfnamefont {P.~O.}\ \bibnamefont {Colard}}, \bibinfo {author} {\bibfnamefont {J.~M.}\ \bibnamefont {Schloss}}, \bibinfo {author} {\bibfnamefont {K.~S.}\ \bibnamefont {Olsson}}, \bibinfo {author} {\bibfnamefont {R.}~\bibnamefont {Trubko}}, \bibinfo {author} {\bibfnamefont {M.~L.}\ \bibnamefont {Markham}}, \bibinfo {author} {\bibfnamefont {A.}~\bibnamefont {Rathmill}}, \bibinfo {author} {\bibfnamefont {B.}~\bibnamefont {Horne-Smith}}, \bibinfo {author} {\bibfnamefont {W.}~\bibnamefont {Lew}}, \bibinfo {author} {\bibfnamefont {A.}~\bibnamefont {Manickam}}, \bibinfo {author} {\bibfnamefont {S.}~\bibnamefont {Bruce}}, \bibinfo {author} {\bibfnamefont {P.~G.}\ \bibnamefont {Kaup}}, \bibinfo {author} {\bibfnamefont {J.~C.}\ \bibnamefont {Russo}}, \bibinfo {author} {\bibfnamefont
  {M.~J.}\ \bibnamefont {DiMario}}, \bibinfo {author} {\bibfnamefont {J.~T.}\ \bibnamefont {South}}, \bibinfo {author} {\bibfnamefont {J.~T.}\ \bibnamefont {Hansen}}, \bibinfo {author} {\bibfnamefont {D.~J.}\ \bibnamefont {Twitchen}},\ and\ \bibinfo {author} {\bibfnamefont {R.~L.}\ \bibnamefont {Walsworth}},\ }\bibfield  {title} {\bibinfo {title} {{Characterisation of CVD diamond with high concentrations of nitrogen for magnetic-field sensing applications}},\ }\href {https://doi.org/10.1088/2633-4356/ABD88A} {\bibfield  {journal} {\bibinfo  {journal} {Materials for Quantum Technology}\ }\textbf {\bibinfo {volume} {1}},\ \bibinfo {pages} {025001} (\bibinfo {year} {2021})}\BibitemShut {NoStop}%
\bibitem [{\citenamefont {Beha}\ \emph {et~al.}(2012)\citenamefont {Beha}, \citenamefont {Batalov}, \citenamefont {Manson}, \citenamefont {Bratschitsch},\ and\ \citenamefont {Leitenstorfer}}]{Beha2012OptimumDiamond}%
  \BibitemOpen
  \bibfield  {author} {\bibinfo {author} {\bibfnamefont {K.}~\bibnamefont {Beha}}, \bibinfo {author} {\bibfnamefont {A.}~\bibnamefont {Batalov}}, \bibinfo {author} {\bibfnamefont {N.~B.}\ \bibnamefont {Manson}}, \bibinfo {author} {\bibfnamefont {R.}~\bibnamefont {Bratschitsch}},\ and\ \bibinfo {author} {\bibfnamefont {A.}~\bibnamefont {Leitenstorfer}},\ }\bibfield  {title} {\bibinfo {title} {{Optimum Photoluminescence Excitation and Recharging Cycle of Single Nitrogen-Vacancy Centers in Ultrapure Diamond}},\ }\href {https://doi.org/10.1103/PhysRevLett.109.097404} {\bibfield  {journal} {\bibinfo  {journal} {Physical Review Letters}\ }\textbf {\bibinfo {volume} {109}},\ \bibinfo {pages} {097404} (\bibinfo {year} {2012})}\BibitemShut {NoStop}%
\bibitem [{\citenamefont {Siyushev}\ \emph {et~al.}(2013)\citenamefont {Siyushev}, \citenamefont {Pinto}, \citenamefont {V{\"{o}}r{\"{o}}s}, \citenamefont {Gali}, \citenamefont {Jelezko},\ and\ \citenamefont {Wrachtrup}}]{Siyushev2013OpticallyTemperatures}%
  \BibitemOpen
  \bibfield  {author} {\bibinfo {author} {\bibfnamefont {P.}~\bibnamefont {Siyushev}}, \bibinfo {author} {\bibfnamefont {H.}~\bibnamefont {Pinto}}, \bibinfo {author} {\bibfnamefont {M.}~\bibnamefont {V{\"{o}}r{\"{o}}s}}, \bibinfo {author} {\bibfnamefont {A.}~\bibnamefont {Gali}}, \bibinfo {author} {\bibfnamefont {F.}~\bibnamefont {Jelezko}},\ and\ \bibinfo {author} {\bibfnamefont {J.}~\bibnamefont {Wrachtrup}},\ }\bibfield  {title} {\bibinfo {title} {{Optically Controlled Switching of the Charge State of a Single Nitrogen-Vacancy Center in Diamond at Cryogenic Temperatures}},\ }\href {https://doi.org/10.1103/PhysRevLett.110.167402} {\bibfield  {journal} {\bibinfo  {journal} {Physical Review Letters}\ }\textbf {\bibinfo {volume} {110}},\ \bibinfo {pages} {167402} (\bibinfo {year} {2013})}\BibitemShut {NoStop}%
\bibitem [{\citenamefont {Razinkovas}\ \emph {et~al.}(2021)\citenamefont {Razinkovas}, \citenamefont {Maciaszek}, \citenamefont {Reinhard}, \citenamefont {Doherty},\ and\ \citenamefont {Alkauskas}}]{Razinkovas2021PhotoionizationCalculations}%
  \BibitemOpen
  \bibfield  {author} {\bibinfo {author} {\bibfnamefont {L.}~\bibnamefont {Razinkovas}}, \bibinfo {author} {\bibfnamefont {M.}~\bibnamefont {Maciaszek}}, \bibinfo {author} {\bibfnamefont {F.}~\bibnamefont {Reinhard}}, \bibinfo {author} {\bibfnamefont {M.~W.}\ \bibnamefont {Doherty}},\ and\ \bibinfo {author} {\bibfnamefont {A.}~\bibnamefont {Alkauskas}},\ }\bibfield  {title} {\bibinfo {title} {{Photoionization of negatively charged NV centers in diamond: Theory and ab initio calculations}},\ }\href {https://doi.org/10.1103/PhysRevB.104.235301} {\bibfield  {journal} {\bibinfo  {journal} {Physical Review B}\ }\textbf {\bibinfo {volume} {104}},\ \bibinfo {pages} {235301} (\bibinfo {year} {2021})}\BibitemShut {NoStop}%
\bibitem [{\citenamefont {Smith}\ \emph {et~al.}(2019)\citenamefont {Smith}, \citenamefont {Meynell}, \citenamefont {Bleszynski~Jayich},\ and\ \citenamefont {Meijer}}]{Smith2019ColourTechnologies}%
  \BibitemOpen
  \bibfield  {author} {\bibinfo {author} {\bibfnamefont {J.~M.}\ \bibnamefont {Smith}}, \bibinfo {author} {\bibfnamefont {S.~A.}\ \bibnamefont {Meynell}}, \bibinfo {author} {\bibfnamefont {A.~C.}\ \bibnamefont {Bleszynski~Jayich}},\ and\ \bibinfo {author} {\bibfnamefont {J.}~\bibnamefont {Meijer}},\ }\bibfield  {title} {\bibinfo {title} {{Colour centre generation in diamond for quantum technologies}},\ }\href {https://doi.org/10.1515/nanoph-2019-0196} {\bibfield  {journal} {\bibinfo  {journal} {Nanophotonics}\ }\textbf {\bibinfo {volume} {8}},\ \bibinfo {pages} {1889} (\bibinfo {year} {2019})}\BibitemShut {NoStop}%
\bibitem [{\citenamefont {Orwa}\ \emph {et~al.}(2012)\citenamefont {Orwa}, \citenamefont {Ganesan}, \citenamefont {Newnham}, \citenamefont {Santori}, \citenamefont {Barclay}, \citenamefont {Fu}, \citenamefont {Beausoleil}, \citenamefont {Aharonovich}, \citenamefont {Fairchild}, \citenamefont {Olivero}, \citenamefont {Greentree},\ and\ \citenamefont {Prawer}}]{Orwa2012AnDiamond}%
  \BibitemOpen
  \bibfield  {author} {\bibinfo {author} {\bibfnamefont {J.~O.}\ \bibnamefont {Orwa}}, \bibinfo {author} {\bibfnamefont {K.}~\bibnamefont {Ganesan}}, \bibinfo {author} {\bibfnamefont {J.}~\bibnamefont {Newnham}}, \bibinfo {author} {\bibfnamefont {C.}~\bibnamefont {Santori}}, \bibinfo {author} {\bibfnamefont {P.}~\bibnamefont {Barclay}}, \bibinfo {author} {\bibfnamefont {K.~M.}\ \bibnamefont {Fu}}, \bibinfo {author} {\bibfnamefont {R.~G.}\ \bibnamefont {Beausoleil}}, \bibinfo {author} {\bibfnamefont {I.}~\bibnamefont {Aharonovich}}, \bibinfo {author} {\bibfnamefont {B.~A.}\ \bibnamefont {Fairchild}}, \bibinfo {author} {\bibfnamefont {P.}~\bibnamefont {Olivero}}, \bibinfo {author} {\bibfnamefont {A.~D.}\ \bibnamefont {Greentree}},\ and\ \bibinfo {author} {\bibfnamefont {S.}~\bibnamefont {Prawer}},\ }\bibfield  {title} {\bibinfo {title} {{An upper limit on the lateral vacancy diffusion length in diamond}},\ }\href {https://doi.org/10.1016/j.diamond.2012.02.009} {\bibfield  {journal} {\bibinfo  {journal} {Diamond
  and Related Materials}\ }\textbf {\bibinfo {volume} {24}},\ \bibinfo {pages} {6} (\bibinfo {year} {2012})}\BibitemShut {NoStop}%
\bibitem [{\citenamefont {Onoda}\ \emph {et~al.}(2017)\citenamefont {Onoda}, \citenamefont {Tatsumi}, \citenamefont {Haruyama}, \citenamefont {Teraji}, \citenamefont {Isoya}, \citenamefont {Kada}, \citenamefont {Ohshima},\ and\ \citenamefont {Hanaizumi}}]{Onoda2017DiffusionDiamond}%
  \BibitemOpen
  \bibfield  {author} {\bibinfo {author} {\bibfnamefont {S.}~\bibnamefont {Onoda}}, \bibinfo {author} {\bibfnamefont {K.}~\bibnamefont {Tatsumi}}, \bibinfo {author} {\bibfnamefont {M.}~\bibnamefont {Haruyama}}, \bibinfo {author} {\bibfnamefont {T.}~\bibnamefont {Teraji}}, \bibinfo {author} {\bibfnamefont {J.}~\bibnamefont {Isoya}}, \bibinfo {author} {\bibfnamefont {W.}~\bibnamefont {Kada}}, \bibinfo {author} {\bibfnamefont {T.}~\bibnamefont {Ohshima}},\ and\ \bibinfo {author} {\bibfnamefont {O.}~\bibnamefont {Hanaizumi}},\ }\bibfield  {title} {\bibinfo {title} {{Diffusion of Vacancies Created by High-Energy Heavy Ion Strike Into Diamond}},\ }\href {https://doi.org/10.1002/pssa.201700160} {\bibfield  {journal} {\bibinfo  {journal} {Physica Status Solidi (A) Applications and Materials Science}\ }\textbf {\bibinfo {volume} {214}},\ \bibinfo {pages} {1700160} (\bibinfo {year} {2017})}\BibitemShut {NoStop}%
\bibitem [{\citenamefont {Steeds}\ and\ \citenamefont {Kohn}(2014)}]{Steeds2014AnnealingSamples}%
  \BibitemOpen
  \bibfield  {author} {\bibinfo {author} {\bibfnamefont {J.~W.}\ \bibnamefont {Steeds}}\ and\ \bibinfo {author} {\bibfnamefont {S.}~\bibnamefont {Kohn}},\ }\bibfield  {title} {\bibinfo {title} {{Annealing of electron radiation damage in a wide range of Ib and IIa diamond samples}},\ }\href {https://doi.org/10.1016/j.diamond.2014.09.012} {\bibfield  {journal} {\bibinfo  {journal} {Diamond and Related Materials}\ }\textbf {\bibinfo {volume} {50}},\ \bibinfo {pages} {110} (\bibinfo {year} {2014})}\BibitemShut {NoStop}%
\bibitem [{\citenamefont {Pezzagna}\ \emph {et~al.}(2011)\citenamefont {Pezzagna}, \citenamefont {Rogalla}, \citenamefont {Wildanger}, \citenamefont {Meijer},\ and\ \citenamefont {Zaitsev}}]{Pezzagna2011CreationRemarks}%
  \BibitemOpen
  \bibfield  {author} {\bibinfo {author} {\bibfnamefont {S.}~\bibnamefont {Pezzagna}}, \bibinfo {author} {\bibfnamefont {D.}~\bibnamefont {Rogalla}}, \bibinfo {author} {\bibfnamefont {D.}~\bibnamefont {Wildanger}}, \bibinfo {author} {\bibfnamefont {J.}~\bibnamefont {Meijer}},\ and\ \bibinfo {author} {\bibfnamefont {A.}~\bibnamefont {Zaitsev}},\ }\bibfield  {title} {\bibinfo {title} {{Creation and nature of optical centres in diamond for single-photon emission-overview and critical remarks}},\ }\href {https://doi.org/10.1088/1367-2630/13/3/035024} {\bibfield  {journal} {\bibinfo  {journal} {New Journal of Physics}\ }\textbf {\bibinfo {volume} {13}},\ \bibinfo {pages} {035024} (\bibinfo {year} {2011})}\BibitemShut {NoStop}%
\bibitem [{\citenamefont {McLellan}\ \emph {et~al.}(2016)\citenamefont {McLellan}, \citenamefont {Myers}, \citenamefont {Kraemer}, \citenamefont {Ohno}, \citenamefont {Awschalom},\ and\ \citenamefont {Bleszynski~Jayich}}]{McLellan2016PatternedTechnique}%
  \BibitemOpen
  \bibfield  {author} {\bibinfo {author} {\bibfnamefont {C.~A.}\ \bibnamefont {McLellan}}, \bibinfo {author} {\bibfnamefont {B.~A.}\ \bibnamefont {Myers}}, \bibinfo {author} {\bibfnamefont {S.}~\bibnamefont {Kraemer}}, \bibinfo {author} {\bibfnamefont {K.}~\bibnamefont {Ohno}}, \bibinfo {author} {\bibfnamefont {D.~D.}\ \bibnamefont {Awschalom}},\ and\ \bibinfo {author} {\bibfnamefont {A.~C.}\ \bibnamefont {Bleszynski~Jayich}},\ }\bibfield  {title} {\bibinfo {title} {{Patterned Formation of Highly Coherent Nitrogen-Vacancy Centers Using a Focused Electron Irradiation Technique}},\ }\href {https://doi.org/10.1021/acs.nanolett.5b05304} {\bibfield  {journal} {\bibinfo  {journal} {Nano Letters}\ }\textbf {\bibinfo {volume} {16}},\ \bibinfo {pages} {2450} (\bibinfo {year} {2016})}\BibitemShut {NoStop}%
\bibitem [{\citenamefont {Chen}\ \emph {et~al.}(2017)\citenamefont {Chen}, \citenamefont {Salter}, \citenamefont {Knauer}, \citenamefont {Weng}, \citenamefont {Frangeskou}, \citenamefont {Stephen}, \citenamefont {Ishmael}, \citenamefont {Dolan}, \citenamefont {Johnson}, \citenamefont {Green}, \citenamefont {Morley}, \citenamefont {Newton}, \citenamefont {Rarity}, \citenamefont {Booth},\ and\ \citenamefont {Smith}}]{Chen2017LaserDiamond}%
  \BibitemOpen
  \bibfield  {author} {\bibinfo {author} {\bibfnamefont {Y.-C.}\ \bibnamefont {Chen}}, \bibinfo {author} {\bibfnamefont {P.~S.}\ \bibnamefont {Salter}}, \bibinfo {author} {\bibfnamefont {S.}~\bibnamefont {Knauer}}, \bibinfo {author} {\bibfnamefont {L.}~\bibnamefont {Weng}}, \bibinfo {author} {\bibfnamefont {A.~C.}\ \bibnamefont {Frangeskou}}, \bibinfo {author} {\bibfnamefont {C.~J.}\ \bibnamefont {Stephen}}, \bibinfo {author} {\bibfnamefont {S.~N.}\ \bibnamefont {Ishmael}}, \bibinfo {author} {\bibfnamefont {P.~R.}\ \bibnamefont {Dolan}}, \bibinfo {author} {\bibfnamefont {S.}~\bibnamefont {Johnson}}, \bibinfo {author} {\bibfnamefont {B.~L.}\ \bibnamefont {Green}}, \bibinfo {author} {\bibfnamefont {G.~W.}\ \bibnamefont {Morley}}, \bibinfo {author} {\bibfnamefont {M.~E.}\ \bibnamefont {Newton}}, \bibinfo {author} {\bibfnamefont {J.~G.}\ \bibnamefont {Rarity}}, \bibinfo {author} {\bibfnamefont {M.~J.}\ \bibnamefont {Booth}},\ and\ \bibinfo {author} {\bibfnamefont {J.~M.}\ \bibnamefont {Smith}},\ }\bibfield
  {title} {\bibinfo {title} {{Laser writing of coherent colour centres in diamond}},\ }\href {https://doi.org/10.1038/NPHOTON.2016.234} {\bibfield  {journal} {\bibinfo  {journal} {Nature Photonics I}\ }\textbf {\bibinfo {volume} {11}},\ \bibinfo {pages} {77} (\bibinfo {year} {2017})}\BibitemShut {NoStop}%
\bibitem [{\citenamefont {Campbell}\ and\ \citenamefont {Mainwood}(2000)}]{Campbell2000RadiationIrradiation}%
  \BibitemOpen
  \bibfield  {author} {\bibinfo {author} {\bibfnamefont {B.}~\bibnamefont {Campbell}}\ and\ \bibinfo {author} {\bibfnamefont {A.}~\bibnamefont {Mainwood}},\ }\bibfield  {title} {\bibinfo {title} {{Radiation damage of diamond by electron and gamma irradiation}},\ }\href@noop {} {\bibfield  {journal} {\bibinfo  {journal} {phys. stat. sol. (a)}\ }\textbf {\bibinfo {volume} {181}},\ \bibinfo {pages} {5} (\bibinfo {year} {2000})}\BibitemShut {NoStop}%
\bibitem [{\citenamefont {Losero}\ \emph {et~al.}(2023)\citenamefont {Losero}, \citenamefont {Goblot}, \citenamefont {Zhu}, \citenamefont {Babashah}, \citenamefont {Boureau}, \citenamefont {Burkart},\ and\ \citenamefont {Galland}}]{Losero2023CreationIrradiation}%
  \BibitemOpen
  \bibfield  {author} {\bibinfo {author} {\bibfnamefont {E.}~\bibnamefont {Losero}}, \bibinfo {author} {\bibfnamefont {V.}~\bibnamefont {Goblot}}, \bibinfo {author} {\bibfnamefont {Y.}~\bibnamefont {Zhu}}, \bibinfo {author} {\bibfnamefont {H.}~\bibnamefont {Babashah}}, \bibinfo {author} {\bibfnamefont {V.}~\bibnamefont {Boureau}}, \bibinfo {author} {\bibfnamefont {F.}~\bibnamefont {Burkart}},\ and\ \bibinfo {author} {\bibfnamefont {C.}~\bibnamefont {Galland}},\ }\bibfield  {title} {\bibinfo {title} {{Creation of NV Centers in Diamond under 155 MeV Electron Irradiation}},\ }\href {https://doi.org/10.1002/apxr.202300071} {\bibfield  {journal} {\bibinfo  {journal} {Advanced Physics Research}\ }\textbf {\bibinfo {volume} {2300071}},\ \bibinfo {pages} {1} (\bibinfo {year} {2023})}\BibitemShut {NoStop}%
\bibitem [{\citenamefont {Wang}\ \emph {et~al.}(2017)\citenamefont {Wang}, \citenamefont {Steeds}, \citenamefont {Li},\ and\ \citenamefont {Wang}}]{Wang2017AnnealingIrradiation}%
  \BibitemOpen
  \bibfield  {author} {\bibinfo {author} {\bibfnamefont {K.}~\bibnamefont {Wang}}, \bibinfo {author} {\bibfnamefont {J.~W.}\ \bibnamefont {Steeds}}, \bibinfo {author} {\bibfnamefont {Z.}~\bibnamefont {Li}},\ and\ \bibinfo {author} {\bibfnamefont {H.}~\bibnamefont {Wang}},\ }\bibfield  {title} {\bibinfo {title} {{Annealing and lateral migration of defects in IIa diamond created by near-threshold electron irradiation}},\ }\href {https://doi.org/10.1063/1.4980017} {\bibfield  {journal} {\bibinfo  {journal} {Applied Physics Letters}\ }\textbf {\bibinfo {volume} {110}},\ \bibinfo {pages} {152101} (\bibinfo {year} {2017})}\BibitemShut {NoStop}%
\bibitem [{\citenamefont {{Element Six Technologies}}(2023)}]{ElementSixTechnologies2023Https://e6cvd.com/us/application/quantum-radiation/el-sc-plate-2-0x2-0x0-5mm.html}%
  \BibitemOpen
  \bibfield  {author} {\bibinfo {author} {\bibnamefont {{Element Six Technologies}}},\ }\href@noop {} {\bibinfo {title} {{https://e6cvd.com/us/application/quantum-radiation/el-sc-plate-2-0x2-0x0-5mm.html}}} (\bibinfo {year} {2023})\BibitemShut {NoStop}%
\bibitem [{\citenamefont {Davies}\ \emph {et~al.}(1992)\citenamefont {Davies}, \citenamefont {Lawson}, \citenamefont {Collins}, \citenamefont {Mainwood},\ and\ \citenamefont {Sharp}}]{Davies1992Vacancy-relatedDiamond}%
  \BibitemOpen
  \bibfield  {author} {\bibinfo {author} {\bibfnamefont {G.}~\bibnamefont {Davies}}, \bibinfo {author} {\bibfnamefont {S.~C.}\ \bibnamefont {Lawson}}, \bibinfo {author} {\bibfnamefont {A.~T.}\ \bibnamefont {Collins}}, \bibinfo {author} {\bibfnamefont {A.}~\bibnamefont {Mainwood}},\ and\ \bibinfo {author} {\bibfnamefont {S.~J.}\ \bibnamefont {Sharp}},\ }\bibfield  {title} {\bibinfo {title} {{Vacancy-related centers in diamond}},\ }\href@noop {} {\bibfield  {journal} {\bibinfo  {journal} {Physical Review B}\ }\textbf {\bibinfo {volume} {46}},\ \bibinfo {pages} {13 157} (\bibinfo {year} {1992})}\BibitemShut {NoStop}%
\bibitem [{\citenamefont {Twitchen}\ \emph {et~al.}(1999)\citenamefont {Twitchen}, \citenamefont {Hunt}, \citenamefont {Smart}, \citenamefont {Newton},\ and\ \citenamefont {Baker}}]{Twitchen1999CorrelationEPR}%
  \BibitemOpen
  \bibfield  {author} {\bibinfo {author} {\bibfnamefont {D.~J.}\ \bibnamefont {Twitchen}}, \bibinfo {author} {\bibfnamefont {D.~C.}\ \bibnamefont {Hunt}}, \bibinfo {author} {\bibfnamefont {V.}~\bibnamefont {Smart}}, \bibinfo {author} {\bibfnamefont {M.~E.}\ \bibnamefont {Newton}},\ and\ \bibinfo {author} {\bibfnamefont {J.~M.}\ \bibnamefont {Baker}},\ }\bibfield  {title} {\bibinfo {title} {{Correlation between ND1 optical absorption and the concentration of negative vacancies determined by electron paramagnetic resonance (EPR)}},\ }\href {www.elsevier.com/locate/diamond} {\bibfield  {journal} {\bibinfo  {journal} {Diamond and Related Materials}\ }\textbf {\bibinfo {volume} {8}},\ \bibinfo {pages} {1572} (\bibinfo {year} {1999})}\BibitemShut {NoStop}%
\bibitem [{\citenamefont {Newsom}(2023)}]{Newsom2023SpectroscopicDiamond}%
  \BibitemOpen
  \bibfield  {author} {\bibinfo {author} {\bibfnamefont {C.}~\bibnamefont {Newsom}},\ }\emph {\bibinfo {title} {{Spectroscopic Characterisation of Defects and Strain in Diamond}}},\ \href@noop {} {Ph.D. thesis},\ \bibinfo  {school} {University of Warwick}, \bibinfo {address} {Coventry} (\bibinfo {year} {2023})\BibitemShut {NoStop}%
\bibitem [{\citenamefont {Kiflawi}\ \emph {et~al.}(1997)\citenamefont {Kiflawi}, \citenamefont {Sittas}, \citenamefont {Kanda},\ and\ \citenamefont {Fisher}}]{Kiflawi1997TheCrystals}%
  \BibitemOpen
  \bibfield  {author} {\bibinfo {author} {\bibfnamefont {I.}~\bibnamefont {Kiflawi}}, \bibinfo {author} {\bibfnamefont {G.}~\bibnamefont {Sittas}}, \bibinfo {author} {\bibfnamefont {H.}~\bibnamefont {Kanda}},\ and\ \bibinfo {author} {\bibfnamefont {D.}~\bibnamefont {Fisher}},\ }\bibfield  {title} {\bibinfo {title} {{The irradiation and annealing of Si-doped diamond single crystals}},\ }\href@noop {} {\bibfield  {journal} {\bibinfo  {journal} {Diamond and Related Materials}\ }\textbf {\bibinfo {volume} {6}},\ \bibinfo {pages} {146} (\bibinfo {year} {1997})}\BibitemShut {NoStop}%
\bibitem [{\citenamefont {Zaitsev}\ \emph {et~al.}(2017)\citenamefont {Zaitsev}, \citenamefont {Moe},\ and\ \citenamefont {Wang}}]{Zaitsev2017OpticalDiamond}%
  \BibitemOpen
  \bibfield  {author} {\bibinfo {author} {\bibfnamefont {A.~M.}\ \bibnamefont {Zaitsev}}, \bibinfo {author} {\bibfnamefont {K.~S.}\ \bibnamefont {Moe}},\ and\ \bibinfo {author} {\bibfnamefont {W.}~\bibnamefont {Wang}},\ }\bibfield  {title} {\bibinfo {title} {{Optical centers and their depth distribution in electron irradiated CVD diamond}},\ }\href {https://doi.org/10.1016/j.diamond.2016.11.015} {\bibfield  {journal} {\bibinfo  {journal} {Diamond and Related Materials}\ }\textbf {\bibinfo {volume} {71}},\ \bibinfo {pages} {38} (\bibinfo {year} {2017})}\BibitemShut {NoStop}%
\bibitem [{\citenamefont {Mills}(2022)}]{Mills2022}%
  \BibitemOpen
  \bibfield  {author} {\bibinfo {author} {\bibfnamefont {L.~A.}\ \bibnamefont {Mills}},\ }\emph {\bibinfo {title} {{Optical Characterisation of Point Defects in CVD Diamond}}},\ \href@noop {} {Ph.D. thesis},\ \bibinfo  {school} {University of Warwick}, \bibinfo {address} {Coventry} (\bibinfo {year} {2022})\BibitemShut {NoStop}%
\bibitem [{\citenamefont {Ruiang Guo Kaiyue~Wang}\ and\ \citenamefont {Wang}(2020)}]{RuiangGuoKaiyueWang2020OpticalDiamond}%
  \BibitemOpen
  \bibfield  {author} {\bibinfo {author} {\bibfnamefont {S.~D.}\ \bibnamefont {Ruiang Guo Kaiyue~Wang}}\ and\ \bibinfo {author} {\bibfnamefont {H.}~\bibnamefont {Wang}},\ }\bibfield  {title} {\bibinfo {title} {{Optical defects and their depth penetration in 200 keV electron irradiated IIa diamond}},\ }\href {https://doi.org/10.1080/10420150.2020.1799376} {\bibfield  {journal} {\bibinfo  {journal} {Radiation Effects and Defects in Solids}\ }\textbf {\bibinfo {volume} {175}},\ \bibinfo {pages} {1083} (\bibinfo {year} {2020})}\BibitemShut {NoStop}%
\bibitem [{\citenamefont {Baker}\ \emph {et~al.}(1997)\citenamefont {Baker}, \citenamefont {Witchen},\ and\ \citenamefont {Newton}}]{Baker1997ElectronIn}%
  \BibitemOpen
  \bibfield  {author} {\bibinfo {author} {\bibfnamefont {J.}~\bibnamefont {Baker}}, \bibinfo {author} {\bibfnamefont {D.}~\bibnamefont {Witchen}},\ and\ \bibinfo {author} {\bibfnamefont {M.}~\bibnamefont {Newton}},\ }\bibfield  {title} {\bibinfo {title} {{Electron paramagnetic resonance data on the defect R1 in}},\ }\href {https://doi.org/10.1080/095008397179381} {\bibfield  {journal} {\bibinfo  {journal} {Philosophical Magazine Letters}\ }\textbf {\bibinfo {volume} {76}},\ \bibinfo {pages} {57} (\bibinfo {year} {1997})}\BibitemShut {NoStop}%
\bibitem [{\citenamefont {Hunt}\ \emph {et~al.}(2000{\natexlab{a}})\citenamefont {Hunt}, \citenamefont {Twitchen}, \citenamefont {Newton}, \citenamefont {Baker}, \citenamefont {Anthony}, \citenamefont {Banholzer},\ and\ \citenamefont {Vagarali}}]{Hunt2000IdentificationDiamond}%
  \BibitemOpen
  \bibfield  {author} {\bibinfo {author} {\bibfnamefont {D.~C.}\ \bibnamefont {Hunt}}, \bibinfo {author} {\bibfnamefont {D.~J.}\ \bibnamefont {Twitchen}}, \bibinfo {author} {\bibfnamefont {M.~E.}\ \bibnamefont {Newton}}, \bibinfo {author} {\bibfnamefont {J.~M.}\ \bibnamefont {Baker}}, \bibinfo {author} {\bibfnamefont {T.~R.}\ \bibnamefont {Anthony}}, \bibinfo {author} {\bibfnamefont {W.~F.}\ \bibnamefont {Banholzer}},\ and\ \bibinfo {author} {\bibfnamefont {S.~S.}\ \bibnamefont {Vagarali}},\ }\bibfield  {title} {\bibinfo {title} {{Identification of the neutral carbon <100>-split interstitial in diamond}},\ }\href@noop {} {\bibfield  {journal} {\bibinfo  {journal} {Physical Review B}\ }\textbf {\bibinfo {volume} {61}},\ \bibinfo {pages} {3863} (\bibinfo {year} {2000}{\natexlab{a}})}\BibitemShut {NoStop}%
\bibitem [{\citenamefont {Hunt}\ \emph {et~al.}(2000{\natexlab{b}})\citenamefont {Hunt}, \citenamefont {Twitchen}, \citenamefont {Newton}, \citenamefont {Baker}, \citenamefont {Kirui}, \citenamefont {van Wyk}, \citenamefont {Anthony},\ and\ \citenamefont {Banholzer}}]{Hunt2000EPRDiamond}%
  \BibitemOpen
  \bibfield  {author} {\bibinfo {author} {\bibfnamefont {D.}~\bibnamefont {Hunt}}, \bibinfo {author} {\bibfnamefont {D.}~\bibnamefont {Twitchen}}, \bibinfo {author} {\bibfnamefont {M.}~\bibnamefont {Newton}}, \bibinfo {author} {\bibfnamefont {J.}~\bibnamefont {Baker}}, \bibinfo {author} {\bibfnamefont {J.}~\bibnamefont {Kirui}}, \bibinfo {author} {\bibfnamefont {J.}~\bibnamefont {van Wyk}}, \bibinfo {author} {\bibfnamefont {T.}~\bibnamefont {Anthony}},\ and\ \bibinfo {author} {\bibfnamefont {W.}~\bibnamefont {Banholzer}},\ }\bibfield  {title} {\bibinfo {title} {{EPR data on the self-interstitial complex O3 in diamond}},\ }\href {https://doi.org/10.1103/PhysRevB.62.6587} {\bibfield  {journal} {\bibinfo  {journal} {Physical Review B}\ }\textbf {\bibinfo {volume} {62}},\ \bibinfo {pages} {6587} (\bibinfo {year} {2000}{\natexlab{b}})}\BibitemShut {NoStop}%
\bibitem [{\citenamefont {Goss}\ \emph {et~al.}(2001)\citenamefont {Goss}, \citenamefont {Coomer}, \citenamefont {Shaw}, \citenamefont {Briddon}, \citenamefont {Rayson},\ and\ \citenamefont {{\"{o}}berg}}]{Goss2001Self-interstitialDiamond}%
  \BibitemOpen
  \bibfield  {author} {\bibinfo {author} {\bibfnamefont {J.~P.}\ \bibnamefont {Goss}}, \bibinfo {author} {\bibfnamefont {B.~J.}\ \bibnamefont {Coomer}}, \bibinfo {author} {\bibfnamefont {T.~D.}\ \bibnamefont {Shaw}}, \bibinfo {author} {\bibfnamefont {P.~R.}\ \bibnamefont {Briddon}}, \bibinfo {author} {\bibfnamefont {M.}~\bibnamefont {Rayson}},\ and\ \bibinfo {author} {\bibfnamefont {S.}~\bibnamefont {{\"{o}}berg}},\ }\bibfield  {title} {\bibinfo {title} {{Self-interstitial aggregation in diamond}},\ }\href {https://doi.org/10.1103/PhysRevB.63.195208} {\bibfield  {journal} {\bibinfo  {journal} {Physical Review B - Condensed Matter and Materials Physics}\ }\textbf {\bibinfo {volume} {63}},\ \bibinfo {pages} {195208} (\bibinfo {year} {2001})}\BibitemShut {NoStop}%
\bibitem [{\citenamefont {Newton}\ \emph {et~al.}(2002)\citenamefont {Newton}, \citenamefont {Campbell}, \citenamefont {Twitchen}, \citenamefont {Baker},\ and\ \citenamefont {Anthony}}]{Newton2002Recombination-enhancedDiamond}%
  \BibitemOpen
  \bibfield  {author} {\bibinfo {author} {\bibfnamefont {M.~E.}\ \bibnamefont {Newton}}, \bibinfo {author} {\bibfnamefont {B.~A.}\ \bibnamefont {Campbell}}, \bibinfo {author} {\bibfnamefont {D.~J.}\ \bibnamefont {Twitchen}}, \bibinfo {author} {\bibfnamefont {J.~M.}\ \bibnamefont {Baker}},\ and\ \bibinfo {author} {\bibfnamefont {T.~R.}\ \bibnamefont {Anthony}},\ }\bibfield  {title} {\bibinfo {title} {{Recombination-enhanced diffusion of self-interstitial atoms and vacancy-interstitial recombination in diamond}},\ }\href@noop {} {\bibfield  {journal} {\bibinfo  {journal} {Diamond and Related Materials}\ }\textbf {\bibinfo {volume} {11}},\ \bibinfo {pages} {618} (\bibinfo {year} {2002})}\BibitemShut {NoStop}%
\bibitem [{\citenamefont {Erasmus}(2014)}]{erasmus2014simulation}%
  \BibitemOpen
  \bibfield  {author} {\bibinfo {author} {\bibfnamefont {N.~R.}\ \bibnamefont {Erasmus}},\ }\emph {\bibinfo {title} {Simulation of Silicon and Diamond detector systems by Geant4 simulation techniques}},\ \href@noop {} {Ph.D. thesis},\ \bibinfo  {school} {University of the Western Cape} (\bibinfo {year} {2014})\BibitemShut {NoStop}%
\bibitem [{\citenamefont {Smith}\ \emph {et~al.}(2004)\citenamefont {Smith}, \citenamefont {Davies}, \citenamefont {Newton},\ and\ \citenamefont {Kanda}}]{Smith2004StructureDiamond}%
  \BibitemOpen
  \bibfield  {author} {\bibinfo {author} {\bibfnamefont {H.~E.}\ \bibnamefont {Smith}}, \bibinfo {author} {\bibfnamefont {G.}~\bibnamefont {Davies}}, \bibinfo {author} {\bibfnamefont {M.~E.}\ \bibnamefont {Newton}},\ and\ \bibinfo {author} {\bibfnamefont {H.}~\bibnamefont {Kanda}},\ }\bibfield  {title} {\bibinfo {title} {{Structure of the self-interstitial in diamond}},\ }\href {https://doi.org/10.1103/PhysRevB.69.045203} {\bibfield  {journal} {\bibinfo  {journal} {Physical Review B - Condensed Matter and Materials Physics}\ }\textbf {\bibinfo {volume} {69}},\ \bibinfo {pages} {045203} (\bibinfo {year} {2004})}\BibitemShut {NoStop}%
\bibitem [{\citenamefont {Kiflawi}\ \emph {et~al.}(2007)\citenamefont {Kiflawi}, \citenamefont {Collins}, \citenamefont {Iakoubovskii},\ and\ \citenamefont {Fisher}}]{Kiflawi2007ElectronDiamond}%
  \BibitemOpen
  \bibfield  {author} {\bibinfo {author} {\bibfnamefont {I.}~\bibnamefont {Kiflawi}}, \bibinfo {author} {\bibfnamefont {A.~T.}\ \bibnamefont {Collins}}, \bibinfo {author} {\bibfnamefont {K.}~\bibnamefont {Iakoubovskii}},\ and\ \bibinfo {author} {\bibfnamefont {D.}~\bibnamefont {Fisher}},\ }\bibfield  {title} {\bibinfo {title} {{Electron irradiation and the formation of vacancy-interstitial pairs in diamond}},\ }\href {https://doi.org/10.1088/0953-8984/19/4/046216} {\bibfield  {journal} {\bibinfo  {journal} {Journal of Physics Condensed Matter}\ }\textbf {\bibinfo {volume} {19}},\ \bibinfo {pages} {046216} (\bibinfo {year} {2007})}\BibitemShut {NoStop}%
\bibitem [{\citenamefont {Lomer}\ and\ \citenamefont {Wild}(1973)}]{Lomer1973ElectronTemperatures}%
  \BibitemOpen
  \bibfield  {author} {\bibinfo {author} {\bibfnamefont {J.~N.}\ \bibnamefont {Lomer}}\ and\ \bibinfo {author} {\bibfnamefont {A.~M.~A.}\ \bibnamefont {Wild}},\ }\bibfield  {title} {\bibinfo {title} {{Electron spin resonance in electron irradiated diamond annealed to high temperatures}},\ }\href {https://doi.org/10.1080/00337577308232595} {\bibfield  {journal} {\bibinfo  {journal} {Radiation Effects}\ }\textbf {\bibinfo {volume} {17}},\ \bibinfo {pages} {37} (\bibinfo {year} {1973})}\BibitemShut {NoStop}%
\bibitem [{\citenamefont {Hounsome}\ \emph {et~al.}(2005)\citenamefont {Hounsome}, \citenamefont {Jones}, \citenamefont {Martineau}, \citenamefont {Shaw}, \citenamefont {Briddon}, \citenamefont {{\"{O}}berg}, \citenamefont {Blumenau},\ and\ \citenamefont {Fujita}}]{Hounsome2005OpticalDiamond}%
  \BibitemOpen
  \bibfield  {author} {\bibinfo {author} {\bibfnamefont {L.~S.}\ \bibnamefont {Hounsome}}, \bibinfo {author} {\bibfnamefont {R.}~\bibnamefont {Jones}}, \bibinfo {author} {\bibfnamefont {P.~M.}\ \bibnamefont {Martineau}}, \bibinfo {author} {\bibfnamefont {M.~J.}\ \bibnamefont {Shaw}}, \bibinfo {author} {\bibfnamefont {P.~R.}\ \bibnamefont {Briddon}}, \bibinfo {author} {\bibfnamefont {S.}~\bibnamefont {{\"{O}}berg}}, \bibinfo {author} {\bibfnamefont {A.~T.}\ \bibnamefont {Blumenau}},\ and\ \bibinfo {author} {\bibfnamefont {N.}~\bibnamefont {Fujita}},\ }\bibfield  {title} {\bibinfo {title} {{Optical properties of vacancy related defects in diamond}},\ }\href {https://doi.org/https://doi.org/10.1002/pssa.200561914} {\bibfield  {journal} {\bibinfo  {journal} {physica status solidi (a)}\ }\textbf {\bibinfo {volume} {202}},\ \bibinfo {pages} {2182} (\bibinfo {year} {2005})}\BibitemShut {NoStop}%
\bibitem [{\citenamefont {Jones}\ \emph {et~al.}(2007)\citenamefont {Jones}, \citenamefont {Hounsome}, \citenamefont {Fujita}, \citenamefont {{\"{O}}berg},\ and\ \citenamefont {Briddon}}]{Jones2007ElectricalDiamond}%
  \BibitemOpen
  \bibfield  {author} {\bibinfo {author} {\bibfnamefont {R.}~\bibnamefont {Jones}}, \bibinfo {author} {\bibfnamefont {L.~S.}\ \bibnamefont {Hounsome}}, \bibinfo {author} {\bibfnamefont {N.}~\bibnamefont {Fujita}}, \bibinfo {author} {\bibfnamefont {S.}~\bibnamefont {{\"{O}}berg}},\ and\ \bibinfo {author} {\bibfnamefont {P.~R.}\ \bibnamefont {Briddon}},\ }\bibfield  {title} {\bibinfo {title} {{Electrical and optical properties of multivacancy centres in diamond}},\ }\href {https://doi.org/https://doi.org/10.1002/pssa.200776311} {\bibfield  {journal} {\bibinfo  {journal} {physica status solidi (a)}\ }\textbf {\bibinfo {volume} {204}},\ \bibinfo {pages} {3059} (\bibinfo {year} {2007})}\BibitemShut {NoStop}%
\bibitem [{\citenamefont {Corbett}\ and\ \citenamefont {Watkins}(1965)}]{Corbett1965ProductionSilicon}%
  \BibitemOpen
  \bibfield  {author} {\bibinfo {author} {\bibfnamefont {J.~W.}\ \bibnamefont {Corbett}}\ and\ \bibinfo {author} {\bibfnamefont {G.~D.}\ \bibnamefont {Watkins}},\ }\bibfield  {title} {\bibinfo {title} {{Production of Divacancies and Vacancies by Electron Irradiation of Silicon}},\ }\href {https://doi.org/10.1103/PhysRev.138.A555} {\bibfield  {journal} {\bibinfo  {journal} {Physical Review}\ }\textbf {\bibinfo {volume} {138}},\ \bibinfo {pages} {A555} (\bibinfo {year} {1965})}\BibitemShut {NoStop}%
\bibitem [{\citenamefont {Baker}\ \emph {et~al.}(1999)\citenamefont {Baker}, \citenamefont {Hunt}, \citenamefont {Newton},\ and\ \citenamefont {Twitchen}}]{Baker1999CentresDiamond}%
  \BibitemOpen
  \bibfield  {author} {\bibinfo {author} {\bibfnamefont {J.~M.}\ \bibnamefont {Baker}}, \bibinfo {author} {\bibfnamefont {D.~C.}\ \bibnamefont {Hunt}}, \bibinfo {author} {\bibfnamefont {M.~E.}\ \bibnamefont {Newton}},\ and\ \bibinfo {author} {\bibfnamefont {D.~J.}\ \bibnamefont {Twitchen}},\ }\bibfield  {title} {\bibinfo {title} {{Centres involving two vacancies in diamond}},\ }\href {https://doi.org/10.1080/10420159908230160} {\bibfield  {journal} {\bibinfo  {journal} {Radiation Effects and Defects in Solids}\ }\textbf {\bibinfo {volume} {149}},\ \bibinfo {pages} {233} (\bibinfo {year} {1999})}\BibitemShut {NoStop}%
\bibitem [{\citenamefont {Ashfold}\ \emph {et~al.}(2020)\citenamefont {Ashfold}, \citenamefont {Goss}, \citenamefont {Green}, \citenamefont {May}, \citenamefont {Newton},\ and\ \citenamefont {Peaker}}]{Ashfold2020NitrogenDiamond}%
  \BibitemOpen
  \bibfield  {author} {\bibinfo {author} {\bibfnamefont {M.~N.}\ \bibnamefont {Ashfold}}, \bibinfo {author} {\bibfnamefont {J.~P.}\ \bibnamefont {Goss}}, \bibinfo {author} {\bibfnamefont {B.~L.}\ \bibnamefont {Green}}, \bibinfo {author} {\bibfnamefont {P.~W.}\ \bibnamefont {May}}, \bibinfo {author} {\bibfnamefont {M.~E.}\ \bibnamefont {Newton}},\ and\ \bibinfo {author} {\bibfnamefont {C.~V.}\ \bibnamefont {Peaker}},\ }\bibfield  {title} {\bibinfo {title} {{Nitrogen in Diamond}},\ }\href {https://doi.org/10.1021/acs.chemrev.9b00518} {\bibfield  {journal} {\bibinfo  {journal} {Chemical Reviews}\ }\textbf {\bibinfo {volume} {120}},\ \bibinfo {pages} {5745} (\bibinfo {year} {2020})}\BibitemShut {NoStop}%
\bibitem [{\citenamefont {Chakravarthi}\ \emph {et~al.}(2020)\citenamefont {Chakravarthi}, \citenamefont {Moore}, \citenamefont {Opsvig}, \citenamefont {Pederson}, \citenamefont {Hunt}, \citenamefont {Ivanov}, \citenamefont {Christen}, \citenamefont {Dunham},\ and\ \citenamefont {Fu}}]{Chakravarthi2020WindowCenters}%
  \BibitemOpen
  \bibfield  {author} {\bibinfo {author} {\bibfnamefont {S.}~\bibnamefont {Chakravarthi}}, \bibinfo {author} {\bibfnamefont {C.}~\bibnamefont {Moore}}, \bibinfo {author} {\bibfnamefont {A.}~\bibnamefont {Opsvig}}, \bibinfo {author} {\bibfnamefont {C.}~\bibnamefont {Pederson}}, \bibinfo {author} {\bibfnamefont {E.}~\bibnamefont {Hunt}}, \bibinfo {author} {\bibfnamefont {A.}~\bibnamefont {Ivanov}}, \bibinfo {author} {\bibfnamefont {I.}~\bibnamefont {Christen}}, \bibinfo {author} {\bibfnamefont {S.}~\bibnamefont {Dunham}},\ and\ \bibinfo {author} {\bibfnamefont {K.-M.~C.}\ \bibnamefont {Fu}},\ }\bibfield  {title} {\bibinfo {title} {{Window into NV center kinetics via repeated annealing and spatial tracking of thousands of individual NV centers}},\ }\href {https://doi.org/10.1103/PhysRevMaterials.4.023402} {\bibfield  {journal} {\bibinfo  {journal} {Physical Review Materials}\ }\textbf {\bibinfo {volume} {4}},\ \bibinfo {pages} {23402} (\bibinfo {year} {2020})}\BibitemShut {NoStop}%
\bibitem [{\citenamefont {Santonocito}\ \emph {et~al.}(2024)\citenamefont {Santonocito}, \citenamefont {Denisenko}, \citenamefont {St{\"{o}}hr}, \citenamefont {Knolle}, \citenamefont {Schreck}, \citenamefont {Markham}, \citenamefont {Isoya},\ and\ \citenamefont {Wrachtrup}}]{Santonocito2024NVModelling}%
  \BibitemOpen
  \bibfield  {author} {\bibinfo {author} {\bibfnamefont {S.}~\bibnamefont {Santonocito}}, \bibinfo {author} {\bibfnamefont {A.}~\bibnamefont {Denisenko}}, \bibinfo {author} {\bibfnamefont {R.}~\bibnamefont {St{\"{o}}hr}}, \bibinfo {author} {\bibfnamefont {W.}~\bibnamefont {Knolle}}, \bibinfo {author} {\bibfnamefont {M.}~\bibnamefont {Schreck}}, \bibinfo {author} {\bibfnamefont {M.}~\bibnamefont {Markham}}, \bibinfo {author} {\bibfnamefont {J.}~\bibnamefont {Isoya}},\ and\ \bibinfo {author} {\bibfnamefont {J.}~\bibnamefont {Wrachtrup}},\ }\bibfield  {title} {\bibinfo {title} {{NV centres by vacancies trapping in irradiated diamond: experiments and modelling}},\ }\href {https://doi.org/10.1088/1367-2630/ad2029} {\bibfield  {journal} {\bibinfo  {journal} {New Journal of Physics}\ }\textbf {\bibinfo {volume} {26}},\ \bibinfo {pages} {013054} (\bibinfo {year} {2024})}\BibitemShut {NoStop}%
\bibitem [{\citenamefont {Kuganathan}\ \emph {et~al.}(2023)\citenamefont {Kuganathan}, \citenamefont {Chroneos},\ and\ \citenamefont {Grimes}}]{Kuganathan2023VacancyDiamond}%
  \BibitemOpen
  \bibfield  {author} {\bibinfo {author} {\bibfnamefont {N.}~\bibnamefont {Kuganathan}}, \bibinfo {author} {\bibfnamefont {A.}~\bibnamefont {Chroneos}},\ and\ \bibinfo {author} {\bibfnamefont {R.~W.}\ \bibnamefont {Grimes}},\ }\bibfield  {title} {\bibinfo {title} {{Vacancy defects in nitrogen doped diamond}},\ }\href {https://doi.org/https://doi.org/10.1016/j.physb.2023.414769} {\bibfield  {journal} {\bibinfo  {journal} {Physica B: Condensed Matter}\ }\textbf {\bibinfo {volume} {655}},\ \bibinfo {pages} {414769} (\bibinfo {year} {2023})}\BibitemShut {NoStop}%
\bibitem [{\citenamefont {Slepetz}\ and\ \citenamefont {Kertesz}(2014)}]{Slepetz2014DivacanciesMechanism}%
  \BibitemOpen
  \bibfield  {author} {\bibinfo {author} {\bibfnamefont {B.}~\bibnamefont {Slepetz}}\ and\ \bibinfo {author} {\bibfnamefont {M.}~\bibnamefont {Kertesz}},\ }\bibfield  {title} {\bibinfo {title} {{Divacancies in diamond: a stepwise formation mechanism}},\ }\href {https://doi.org/10.1039/C3CP53384K} {\bibfield  {journal} {\bibinfo  {journal} {Physical Chemistry Chemical Physics}\ }\textbf {\bibinfo {volume} {16}},\ \bibinfo {pages} {1515} (\bibinfo {year} {2014})}\BibitemShut {NoStop}%
\end{thebibliography}%

\end{document}